\UseRawInputEncoding
\RequirePackage{fix-cm}
\documentclass[smallextended]{svjour3}       
\smartqed  
\usepackage{graphicx}
\usepackage{caption}
\usepackage{hyperref}
\usepackage[symbol]{footmisc}
\usepackage{epstopdf}
\usepackage{float}
\usepackage{longtable}
\usepackage{rotating}
\usepackage{amsmath}
\usepackage{pdflscape}
\usepackage{xcolor}
\usepackage{algorithm}
\usepackage{mathptmx}
\usepackage{latexsym}
\usepackage{graphicx}
\usepackage{afterpage}
\usepackage{amsmath}
\usepackage{alphalph}
\usepackage{etoolbox}
\usepackage{rotfloat}
\usepackage{rotating}
\usepackage{multirow}
\usepackage{multicol}
\usepackage{latexsym}
\usepackage{tabularx}
\usepackage{lscape}
\usepackage{amssymb}
\usepackage{algpseudocode}
\usepackage{algorithm}
\usepackage[algo2e]{algorithm2e}
\usepackage{tikz}
\usepackage{pgfplots}
\usepackage{multirow}
\usepackage{pgfplots}
\usepackage{subfig}
\usepackage{geometry}
\usepackage{mathtools}
\usepackage{epstopdf}
\usepackage[symbol]{footmisc}
\def\correspondingauthor{\footnote{Corresponding author.}}
\begin{document}

\title{Towards event aggregation for reducing the volume of logged events during IKC stages of APT attacks
}


\author{Ali Ahmadian Ramaki \and  Abbas Ghaemi-Bafghi\correspondingauthor{}  \and  Abbas Rasoolzadegan
}


\institute{Ali Ahmadian Ramaki \at
              Data and Communication Security Lab, Ferdowsi University of Mashhad, Mashhad, Iran. \\
              \email{ali.ahmadianramaki@mail.um.ac.ir}           
           \and
           Abbas Ghaemi-Bafghi \at
             Data and Communication Security Lab, Ferdowsi University of Mashhad, Mashhad, Iran.\\
              Tel.: +98-51-38805062\\
              Fax: +98-51-38807801\\
              \email{ghaemib@um.ac.ir}           
           \and
           Abbas Rasoolzadegan \at
              Software Quality Lab, Ferdowsi University of Mashhad, Mashhad, Iran.\\
              \email{rasoolzadegan@um.ac.ir}           
}

\vspace{-1cm}
\date{Received: date / Accepted: date}

\maketitle

\begin{abstract}
Nowadays, targeted attacks like Advanced Persistent Threats (APTs) has become one of the major concern of many enterprise networks. As a common approach to counter these attacks, security staff deploy a variety of security and non-security sensors at different lines of defense (Network, Host, and Application) to track the attacker's behaviors during their kill chain. However, one of the drawbacks of this approach is the huge amount of events raised by heterogeneous security and non-security sensors which makes it difficult to analyze logged events for later processing i.e. event correlation for timely detection of APT attacks. Till now, some research papers have been published on event aggregation for reducing the volume of logged low-level events. However, most research works have been provided a method to aggregate the events of a single-type and homogeneous event source i.e. NIDS. In addition, their main focus is only on the degree to which the event volume is reduced, while the amount of security information lost during the event aggregation process is also very important. In this paper, we propose a three-phase event aggregation method to reduce the volume of logged heterogeneous events during APT attacks considering the lowest rate of loss of security information. To this aim, at first, low-level events of the sensors are clustered into some similar event groups and then, after filtering noisy event clusters, the remained clusters are summarized based on an Attribute-Oriented Induction (AOI) method in a controllable manner to reduce the unimportant or duplicated events. The method has been evaluated on the three publicly available datasets: SotM34, Bryant, and LANL. The experimental results show that the method is efficient enough in event aggregation and can reduce events volume up to 99.7\% with an acceptable level of information loss ratio (ILR).
\end{abstract}

\section{Introduction}
\label{1.intro}
Today, the variety of services that can be provided by organizations in the context of computer systems and networks has become a necessity and inevitable. This will make the networks bigger and more sophisticated the technologies used in it i.e. cloud computing, internet of things (IoT), and blockchain. One of the main concerns of networks growth is the occurrence of complex security-related incidents due to increasing the size of the network, poor network management and less hardening of networks against cybersecurity attacks \cite{r1}\cite{r2}. One of the main challenges in large networks is combating a modern type of attack which is getting more and more complex and targeted i.e. advanced persistent threats (APTs) \cite{r3}\cite{r4}. In this type of attack, the attacker, based on a variety of attack tools and techniques, goes through a next-generation threat life cycle to achieve its target, which is called intrusion kill chain (IKC) or cyber kill chain (CKC)\cite{r3}\cite{r5}\cite{r6}\cite{r9}\cite{r10}. In the other words, based on a specified IKC model, the attackers use sequencing of attack stages involving various attack steps to infiltrate the target network and obtain their final goals by moving laterally across multiple hosts of the network for exfiltration of sensitive data or sabotage.

Generally, in an IKC model, the malicious intruders use combined attacks that exploit different vulnerabilities of the targeted network during the attack scenario. In other words, one of the main characteristics of an IKC-based attack is multi-level intrusion which referred to the combined nature of these types of attack that uses all the Network, Host, and Application levels security breaches to perform the attack \cite{r7}\cite{r8}\cite{r9}. Hence, finding the root cause of security incidents will be a challenging task for network security administrators. Although the attackers use advanced attack vectors at the side of the attack surface, some promising approaches are proposed by the security research communities to combat and minimize the attacker's behaviors.

Due to both combined nature of the complex attacks and lack of any well-known attack signatures for the modern IKC-based cybersecurity attacks, traditional network-level security solutions like Firewalls and Intrusion detection systems (IDSs) are not enough to prevent APT attack strategies and block them, solely \cite{r7}\cite{r8}\cite{r9}\cite{r11}. One of the main promising approaches to track the attacker's behaviors and detect the IKC is the use of multiple and various heterogeneous security and non-security sensors in different lines of defense of a monitored network (Network, Host, and Application) to enhance the security level and mitigate the potential risks caused by security breaches \cite{r3}\cite{r7}\cite{r8}\cite{r11}. Security sensors are those that generate security events with high risk and non-security sensors generate ordinary events of the system with low risk.

In an organization with heterogeneous sensors or even total facilities like security operation center (SOC) and security information and event management (SIEM), each sensor in a defense line tends to operate differently on each class of attack regarding intrusion activities of a unique stage of IKC \cite{r8}. Therefore, they record different intrusion reports and log messages in various formats. Although heterogeneous sensors play a vital role to enhance the security of computer networks, the use of them in organizations will bring some main challenges which have been reported in the previous works \cite{r11}\cite{r13}\cite{r15}. One of the main drawbacks of using heterogeneous sensors is the huge amount of security and non-security events/alerts raised by heterogeneous sensors which most of them, about 99\%, are duplicated and have the same root cause or irrelevant and false-positive \cite{r13}\cite{r15}. So, this makes it difficult to analyze logged events for later processing i.e. correlation analysis for complex attack scenario detection like APT attacks. 

Generally, in an event correlation system, the event flooding problem and false-positive events detection is solved by an event aggregation component \cite{r13}\cite{r14}. This component is desirable fusing similar events to an aggregated event (meta-event) and also eliminating the redundant events \cite{r13}\cite{r15}. The main idea of the event aggregation component is reducing the volume of logged events by various sensors during IKC of APT attacks to better managing them for later usages like event correlation analysis that aims to correlate event logs for identifying multi-stage attack scenarios. To the best of our knowledge, there is no suitable IKC-based aggregation method in the presence of heterogeneous sensors in the literature, meanwhile, the use of heterogeneous event aggregation is necessary for timely detection and response to the APT attack. Also, most methods in the literature rely only on aggregating the events of a unique detection level i.e. NIDS sensor from the network level. In addition, most of them try to provide the maximum amount of reduction in the volume of logged events, while they do not care about the lack of security information in the event logs.

In this paper, to overcome the current limitations of the existing event aggregation methods, a heterogeneous event aggregation method is proposed for reducing the volume of logged events volume during the IKC stages of APT attacks. To this main goal, after collecting logged events by the various heterogeneous security and non-security sensors, at first, they are normalized into a common log format to perform other operations related to event aggregation on them. Then, in the first phase, according to a defined time window for receiving events by the aggregation component, they are classified regarding the type of sensor. Afterward, the related events to each sensor are clustered based on an attribute-based similarity matching. Next, in the second phase, the generated event clusters are analyzed based on a clustering-based local outlier factor method to remove and discard noisy events and false-positive. Finally, in the third phase, the resulted events in event clusters are summarized based on an attribute-oriented induction method to summarize events flexibly and form final aggregated events. Based on the experiments on some publicly available datasets namely, SotM34, Bryant, and LANL, the proposed method is capable of reducing the volume of logged events by the heterogeneous sensors in a maximum level with an acceptable level of security information loss. The main contributions of this paper are as follows:
\begin{itemize}
	\item First, we provide a comprehensive mapping between the different IKC stages of complex attacks like APTs and different heterogeneous sensors of the three detection levels (Network, Host, and Application) with their related event features.
	\item Second, we propose a three-phase aggregation method by using clustering-based and attribute-oriented induction methods for aggregating and summarizing heterogeneous events.
	\item Third, we conduct an extensive performance study based on the three publicly available datasets to show the effectiveness of
our approach and compare the method with existing approaches regarding the main standard evaluation metrics.
\end{itemize}

The rest of this paper is organized as follows. In Section~\ref{2:rw}, the relevant literature is reviewed to give background information about the IKC models of APT attacks and the existing research works which have been proposed techniques for event/alert aggregation. Section~\ref{3.prop} presents the proposed event aggregation method with the related algorithms. Also, a concise and useful running example is provided for demonstrating the functions of the proposed method components throughout the section. In Section~\ref{4:eval}, we conduct several experiments based on our proposed method using different well-known and standard datasets in the intrusion detection field for evaluating the correctness and performance of our approach with comparing with the other related works. In addition, in this section, several discussions are presented to analyze the effectiveness of the parameters of the proposed method on each other along with the benefits and advantages of the methods for researchers and practitioners. Finally, Section~\ref{5:conc} concludes the paper and presents some future work.

\section{Literature Review}\label{2:rw}
According to the main goal of this paper mentioned in Section~\ref{1.intro}, it is required to review the existing works for answering the four main research questions (RQs) to obtain this goal and guide us to propose a novel event aggregation method. These RQs are depicted in Table~\ref{tbl:0} along with their rationality to justify their relation to the main goal of this study. In the rest of this section, firstly, the literature is reviewed in Subsection~\ref{2.1.ikc} to Subsection~\ref{2.4.agg} and finally, the results are discussed in Subsection ~\ref{2.5.disc}.

\begin{table}[ht]
\centering
\captionsetup{justification=centering}
\caption{The main research questions (RQs) for the literature review} \label{tbl:0}
\vspace*{-4mm}
\includegraphics[width=1\textwidth]{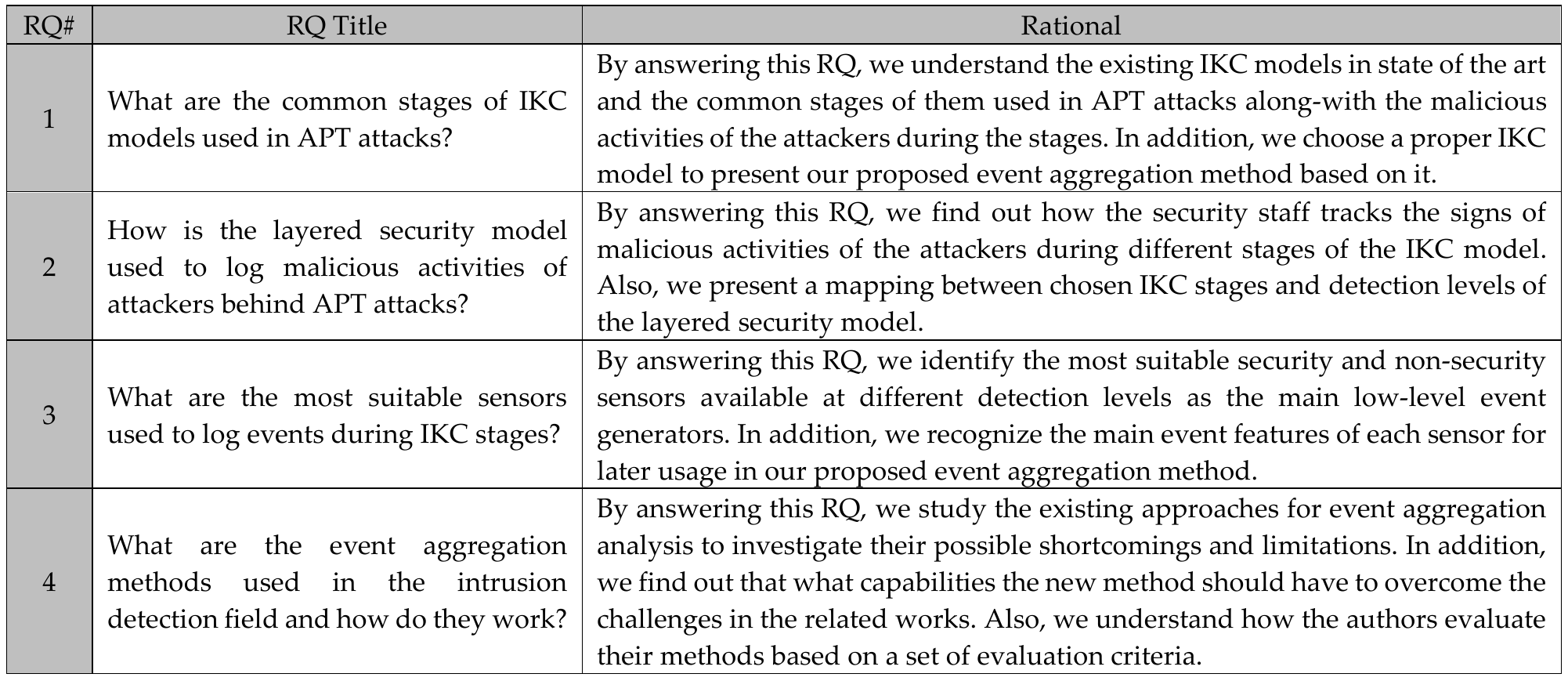}
\end{table}

\vspace{-1cm}
\subsection{Common Stages of IKC Models used in APT Attacks (Regarding the RQ1)}\label{2.1.ikc}
\begin{figure*}[ht]
\captionsetup{justification=centering}
\centering
  \includegraphics[width=0.9\textwidth]{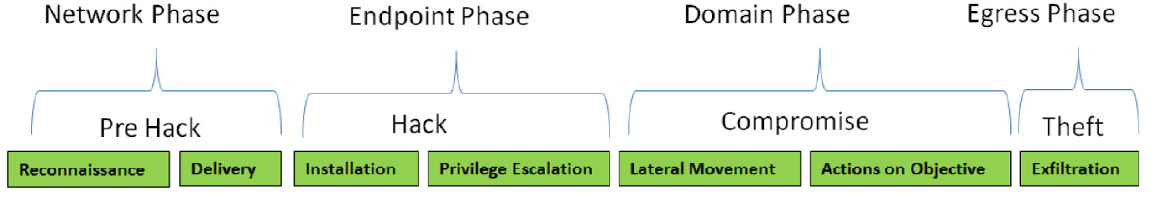}
\caption{Bryant kill chain model macro-phases and attack stages\cite{r9}}
\label{fig:2-1}       
\end{figure*}

To the best of our knowledge, based on a systematic literature review (SLR) method in the area of APT attacks, there are 18 different IKC models proposed in this field both in academic research works including \cite{r7}, \cite{r9}, \cite{r17}, \cite{r18}, \cite{r19}, \cite{r20}, \cite{r21}, and \cite{r22} and reports of industrial companies containing \cite{r23}, \cite{r24}, \cite{r25}, \cite{r26}, \cite{r27}, \cite{r28}, \cite{r29}, and \cite{r30}. Each of all the existing IKC models has a specified number of stages (phases) that are used to perform the targeted attacks by using a set of intrusion activities \cite{r5}. Based on the analysis done in \cite{r9}, most of the introduced models have a common set of attack stages used in APT attacks which has led to the introduction of the Bryant kill chain model. The developed IKC model by Bryant and Saiedian has been made using some modifications to previous models i.e. Lockheed Martin \cite{r9} and Mandiant models \cite{r17} which makes it a suitable choice for data-driven analysis i.e. event aggregation in the security solutions like SIEM, SOC, and threat hunting. The main macro-phases and stages of the Bryant IKC model are depicted in Fig.~\ref{fig:2-1}. 

As shown in Fig.~\ref{fig:2-1}, the Bryant IKC consists of seven stages which are described as follows (abbreviations are used throughout the following sections):

\begin{itemize}
\item \textbf{Reconnaissance (REC)}: In this stage, the attackers gather information about the target. The main objective of this stage is gaining knowledge about the victim and finding the effective methods and technologies to intrude the network.
\item \textbf{Delivery (DEL)}: In this stage, the attackers try to deliver a prepared attack payload to the network by using the established connections between attacker-side nodes to a victim or a collection of victims on the target side.
\item \textbf{Installation (INS)}: In this stage, the main objective of the attackers is the installation of attack payload/malware on the infected systems. This stage is necessary for persistent access in further intrusion activities.
\item \textbf{Privilege Escalation (ESC)}: In this stage, the objective of attackers is to escalate their privileges on the victims which helps them to move the target to find valuable information.
\item \textbf{Lateral Movement (LAT)}: The main objective of this stage, sometimes called internal reconnaissance, is to differentiate between the reconnaissance activity from the external and internal network. This stage is an optional stage based on the attack goals that are used for moving laterally within the compromised network.
\item \textbf{Actions on Objects (ACT)}: By using this stage, the adversaries achieve the attack goals by performing several destructive activities inside the target network which can take months.
\item \textbf{Exfiltration (EXF)}: In this stage, after the successful execution of attack stages on the final objects, the attackers attempt to exfiltrate sensitive data from the target network by using suitable communication and control (C\&C) servers. In addition, to complete the mission, he/she tries to delete intrusion evidence from compromised machines. 
\end{itemize}

Regarding Fig.~\ref{fig:2-1}, it should be noted that some attack stages of the Bryant IKC model are optional and may not be performed by the attackers, or someones may be repeated several times with different conditions by them. In addition, sometimes the stages of the IKC may not be by the order shown in the figure exactly because the attacker may not be successful at one stage and may have to repeat the previous stages of the IKC model. Generally, each stage of the IKC model has some common intrusion activities as attack steps which are used by the attackers for attack progression during the IKC. Table~\ref{tbl:1-1} presents a mapping between the Bryant IKC stages and the commonly used intrusion activities by the adversaries behind APT attacks. Based on the results of this table, the malicious activities used by the attackers range from passive attacks i.e. various types of network scanning, social engineering attacks to active ones i.e. vulnerability exploitation, malicious code execution, software modification, and eventually system destruction or data theft.

\begin{table}[ht]
\centering
\captionsetup{justification=centering}
\caption{Common intrusion activities during each stage of the Bryant IKC model} \label{tbl:1-1}
\vspace*{-4mm}
\includegraphics[width=1\textwidth]{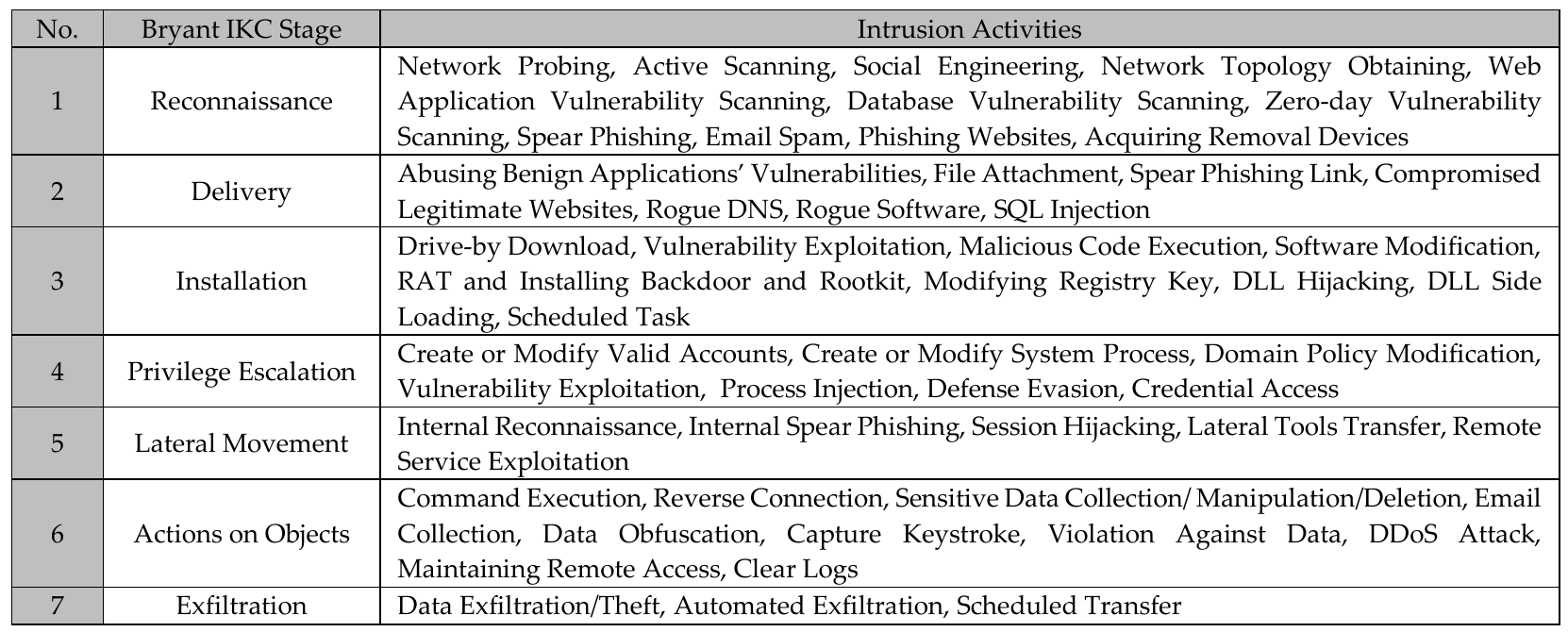}
\end{table} 

\vspace{-1cm}
\subsection{The Role of Layered Security Model for Tracking APT Attacks (Regarding the RQ2)}\label{2.2.lay}
\begin{figure*}
\captionsetup{justification=centering}
\centering
  \includegraphics[width=0.8\textwidth]{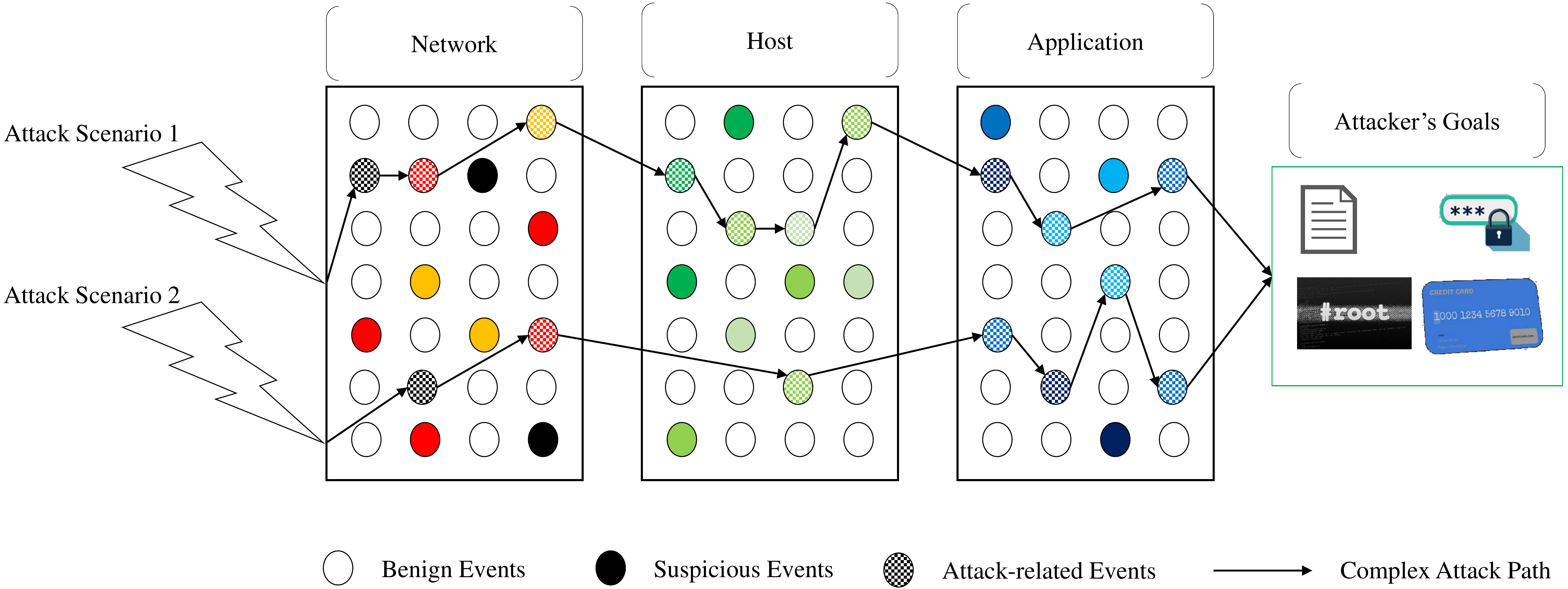}
\caption{Logging events by heterogeneous sensors at different detection levels, Network, Host, and Application}
\label{fig:3}       
\end{figure*}

\begin{figure*}
\captionsetup{justification=centering}
\centering
  \includegraphics[width=0.6\textwidth]{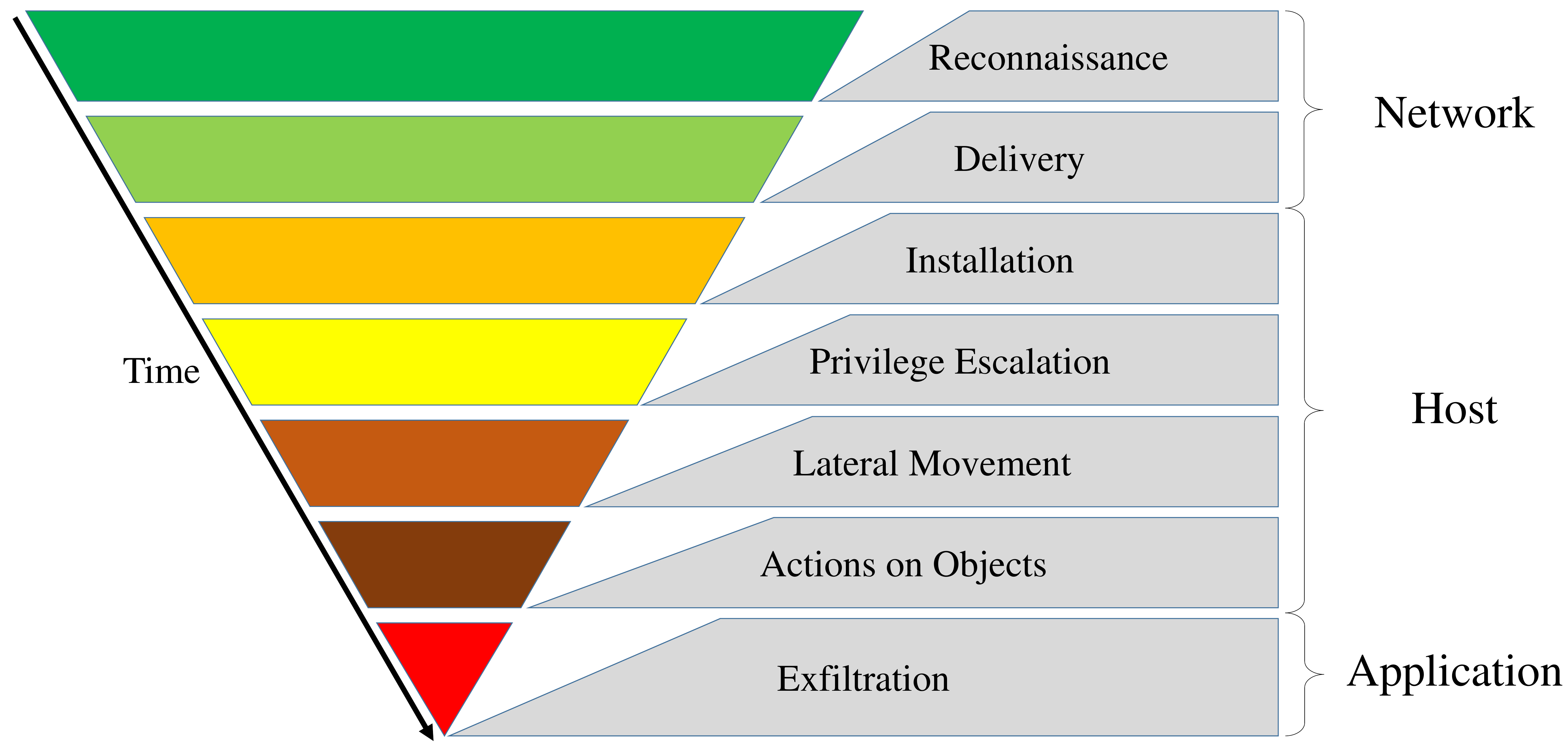}
\caption{Bryant IKC stages mapping to the detection levels, Network, Host, and Application}
\label{fig:2}       
\end{figure*}

In the network security area, the layered security model is defined as the concept of securing a computer network through a sequence of defensive mechanisms, so that if one fails, another is already in place to prevent an attack \cite{r6}\cite{r9}\cite{r34}. The basic assumption on this model is that no single mechanism can successfully safeguard the network from attacks due to the great variety of attacks. According to Fig.~\ref{fig:2-1}, when the attacker initiates an APT attack by using different intrusion activities at the network level, he/she accesses his/her goal in the compromised machines of the internal domain followed by some intrusion activities in the host and application levels \cite{r7}\cite{r8}\cite{r9}. Hence, regarding Fig.~\ref{fig:3}, three main detection levels are used for deploying various security and non-security sensors to log events during the intrusion activities of each IKC stage. In other words, an APT attack scenario based on the IKC model consists of a succession of events logged by both security and non-security sensors. These logged events by heterogeneous sensors with different functionalities can be categorized into three main categories, 1) benign or normal events (uncolored points), 2) suspicious events (colored points), and 3) attack-related events (colored and patterned points). According to Fig.~\ref{fig:3}, 1) a significant part of the logged events during attack scenarios are normal, 2) the variety of logged suspicious events depends on the number of deployed sensors in the monitored network, 3) suspicious events are related to failed attack attempts, and 4) the attack-related events may construct different attack scenarios. A mapping between the IKC stages and the above-mentioned three detection levels is depicted in Fig.~\ref{fig:2}. The colors in this figure, from green to red, indicate the risk level of the attacker's activities advancing the IKC model from low to high. 

In this paper, the selected Bryant IKC model is used for two main purposes, 1) designing a mapping between IKC stages and their related sensors and 2) presenting behavior of APT attackers based on the IKC stages during a sample attack scenario. The first is explained in Section~\ref{3.1.aggr} and the second is provided in the following. To this aim, Fig.~\ref{fig:7} shows an example IKC-based attack scenario and its stages in the form of a sequence diagram to model the attack life cycle. As illustrated in the figure, during the attack scenario, the attacker aims to intrude into the internal network to steal a credential file of the organization. The main steps of the APT attack scenario with the related IKC stages are provided in the following. It should be noted that this example is used in the following sections to understand how the proposed aggregation method works. 

\begin{enumerate}
\item \textbf{Active Scanning (REC)}: The attacker initiates (with IP address 172.25.110.11) the attack by scanning the target network (IP address range 192.168.1.1/24) to find some organization's hosts, which are up for delivering various services.
\item \textbf{Application Vulnerability Scanning (REC)}: After scanning activities and identifying open ports on a specified host (with IP address 192.168.1.12), the attacker attempts to gather information about the system by enumeration. In this step, the attacker finds a vulnerable Adobe Reader application on the enumerated host.
\item \textbf{File Attachment (DEL)}: According to the founded vulnerability, the attacker develops malware, attaches it to a PDF file, and sends the malicious PDF to the target host via an email.
\item \textbf{Drive by Download (INS)}: After opening the Gmail in the browser by the user on the behalf of the targeted host, he/she is tempted to download the attached malicious PDF with an embedded malicious payload.
\item \textbf{Malicious Code Execution (INS)}: After downloading, the user starts the trojaned PDF file installation process. This causes the embedded malicious payload to open a reverse connection to a remote C\&C server.
\item \textbf{Command Execution (ACT)}: After establishing the connection between a compromised host and the C\&C server, the C\&C server sends a command to search for a credential file containing an employee's private information.
\item \textbf{Maintaining Remote Access (ACT)}: The C\&C server forwards some commands to download a vulnerability scanner tool from the server through the file transfer protocol (FTP).
\item \textbf{Sensitive Data Collection (ACT)}: After downloading the scanner, the compromised host as a C\&C client receives commands to run the downloaded scanner and send the credential file to a specified remote web server.
\item \textbf{Data Exfiltration (EXF)}: After executing commands by the infected system, it sends the credential file to the webserver through hypertext transfer protocol (HTTP) protocol.
\item \textbf{Internal Reconnaissance (LAT)}: To collect more information about the targeted network, the attacker starts to move laterally in the internal network by using the conquered machine as a pivot for the later intrusion activities.
\end{enumerate}

\begin{landscape}
\vspace*{1mm}
\begin{figure*}[ht]
\captionsetup{justification=centering}
\centering
  \includegraphics[width=1.2\textwidth,height=0.8\textheight]{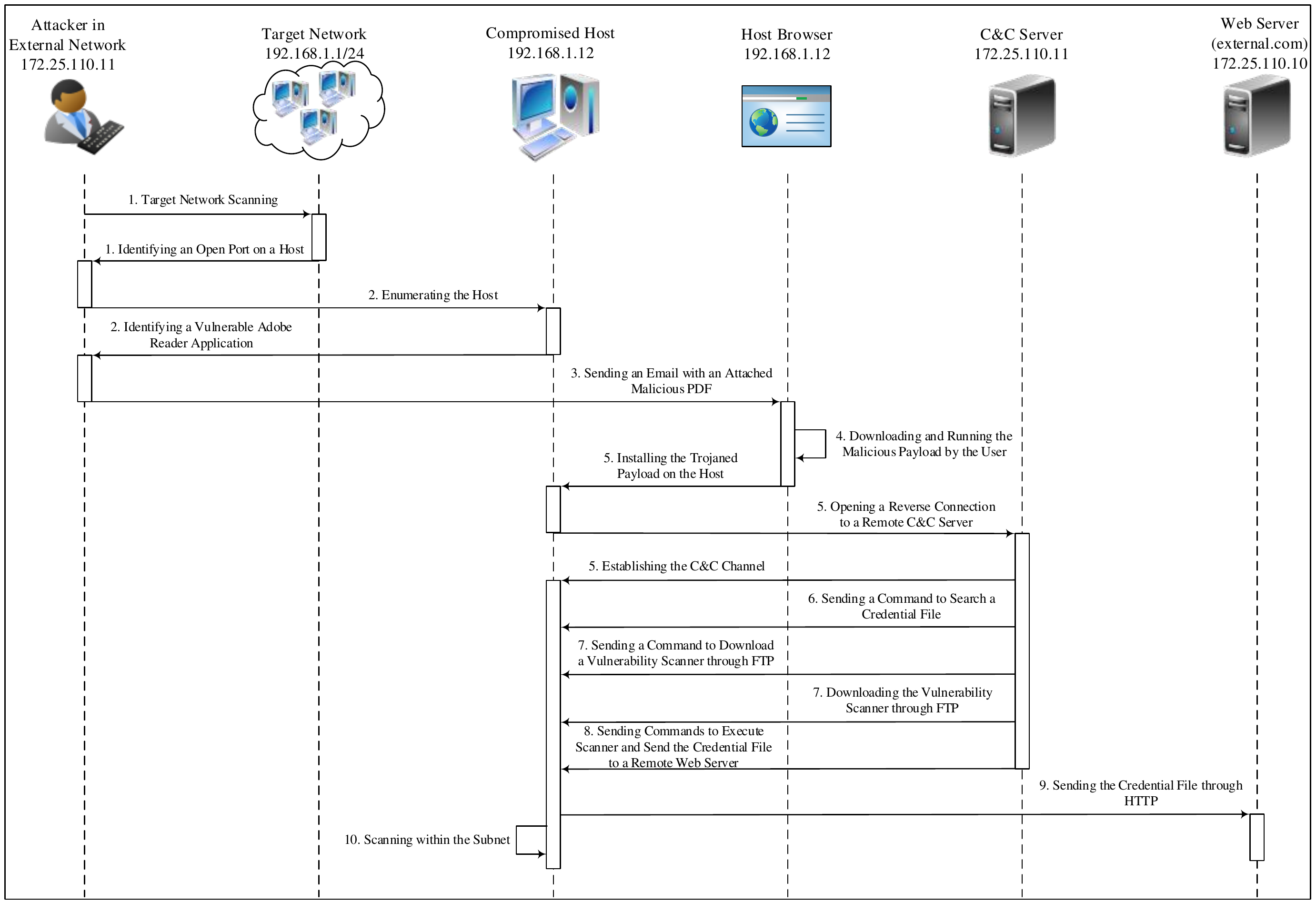}
\caption{A sample APT attack scenario based on the Bryant IKC stages}
\label{fig:7}       
\end{figure*}
\end{landscape}

\subsection{Suitable Sensors for Logging Events during IKC Stages (Regarding the RQ3)}\label{2.3.sen}
\begin{figure*}[h]
\captionsetup{justification=centering}
\centering
  \includegraphics[width=\textwidth]{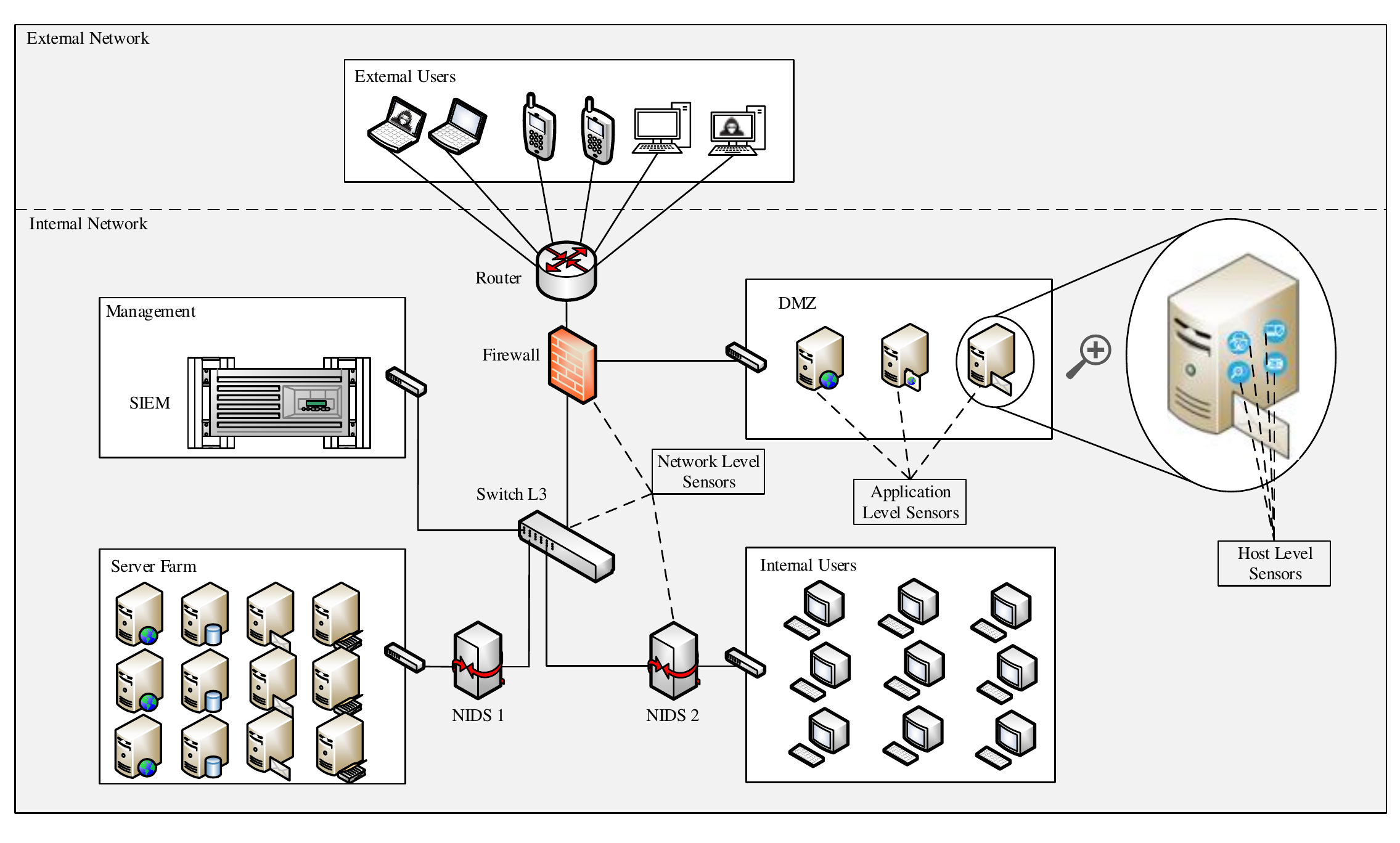}
\caption{Sample monitored network with deployed heterogeneous sensors}
\label{fig:41}       
\end{figure*}

According to the related studies \cite{r5}\cite{r7}\cite{r9}\cite{r12}\cite{r15}\cite{r35}, to track APT attacks, at first, it is required to collect events logged by the heterogeneous sensors which are deployed in different detection levels. Sample monitored network with heterogeneous sensors from different detection levels is shown in Fig.~\ref{fig:41}. As shown in this figure, logged events by the Network level sensors (i.e. IDS, Router, and Switch), Host level sensors (i.e. HIDS, Antivirus, and OS logs), and Application level sensors (Email, Web and Database servers) are transferred by the collector agents to a security incident management platform of the organization like a SIEM solution.

Each sensor has a specified set of attributes/features according to the sensor functionality within a predefined format. Based on our analysis, many studies have been mentioned the main suitable sensors for logging security and non-security events alongside their features, although these are not discussed in details \cite{r8}-\cite{r12}\cite{r31}-\cite{r34}\cite{r36}-\cite{r38}. Table~\ref{tbl:3} presents the most suitable heterogeneous sensors in each detection level with their event features used as the input for the proposed event aggregation method.

As mentioned before, during each attack stage of an IKC model, some low-level events are logged by the various heterogeneous sensors of the different detection levels. For example, a list of logged low-level events during the attack steps of the sample APT attack scenario (Fig~\ref{fig:7}) is shown in Table~\ref{tbl:4}. According to the table, each of the events has a unique sensor-ID and a set of features that has been logged by a sensor with a unique SID provided in Table~\ref{tbl:3}. The empty cells in the table show that the related sensor has no value for the feature. It should be noted that for simplicity, only some important features of the sensors are mentioned in the table.


\begin{landscape}
\begin{table}[ht]
\centering
\captionsetup{justification=centering}
\caption{Heterogeneous security and non-security sensors of different detection levels and their event features\label{tbl:3}}
\vspace*{-4mm}
\includegraphics[width=1.3\textwidth,height=1\textheight]{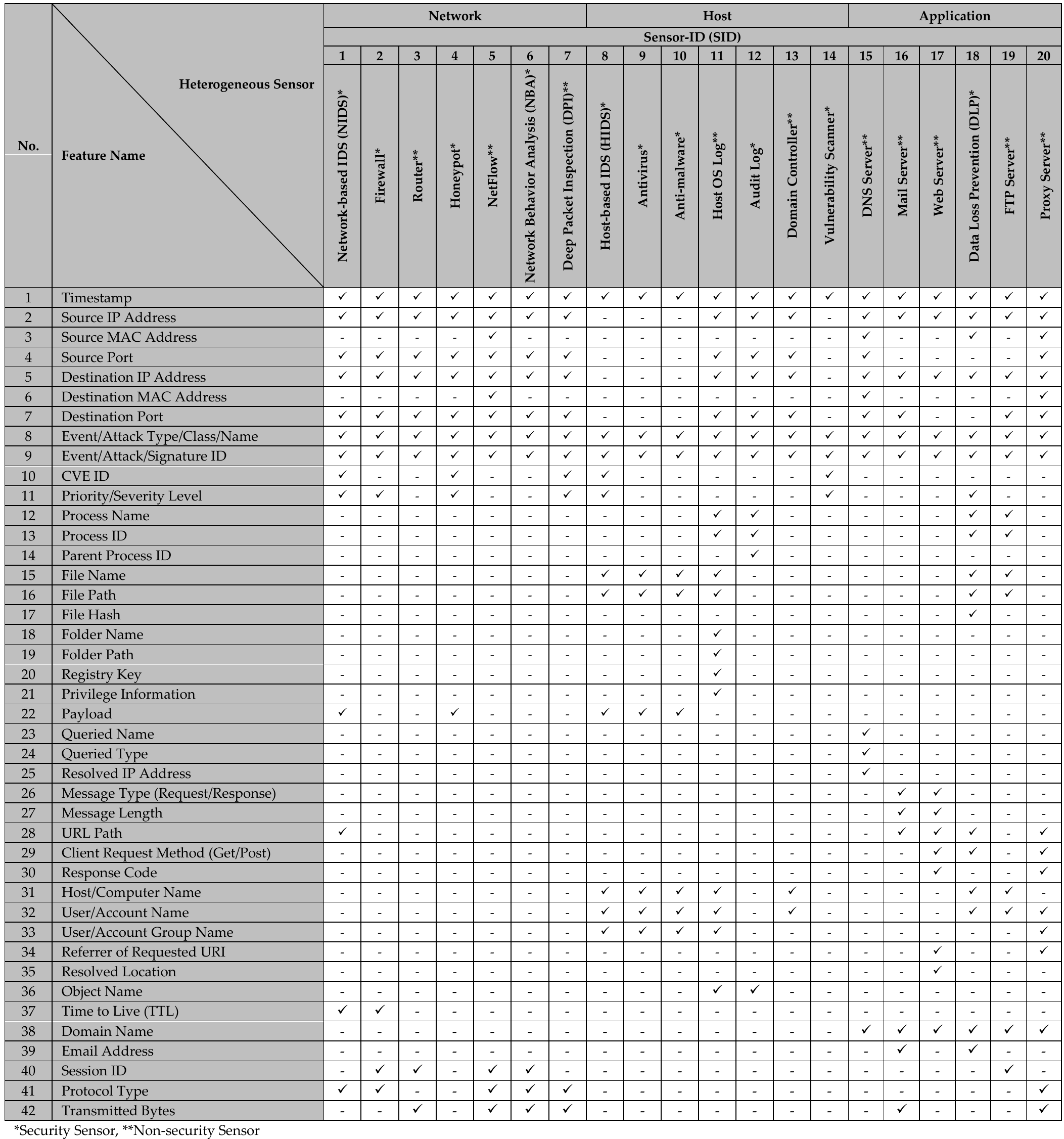}
\end{table}
\end{landscape}

\begin{landscape}
\vspace*{3mm}
\begin{table}[ht]
\centering
\captionsetup{justification=centering}
\caption{Sample logged events by heterogeneous sensors during the sample attack scenario in Fig.~\ref{fig:7}} \label{tbl:4}
\vspace*{-4mm}
\includegraphics[width=1.5\textwidth,height=0.8\textheight]{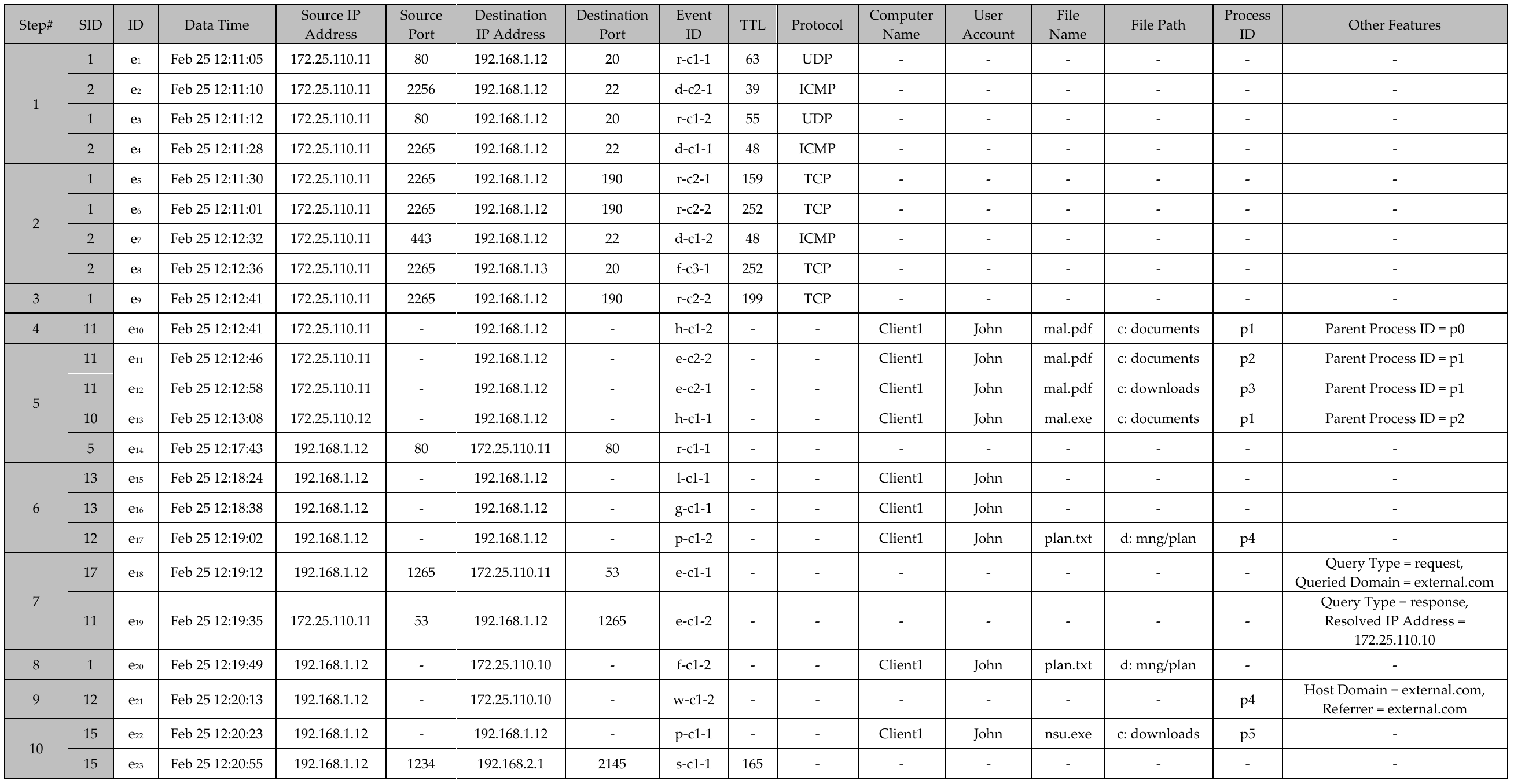}
\end{table}
\end{landscape}

\subsection{The Event Aggregation Methods in the Intrusion Detection Field (Regarding the RQ4)}\label{2.4.agg}
In the literature, there are a lot of research works have been found with different event aggregation approaches for reducing the volume of logged events/alerts of various sensors to improve their quality. Table~\ref{tbl:1} provides a comparative analysis for the existing research works on events/alerts aggregation methods. In the rest of this section, a brief description is provided for the main existing research works in this area.

\begin{table}[ht]
\centering
\captionsetup{justification=centering}
\caption{Overview of the related works to the event aggregation} \label{tbl:1}
\vspace*{-4mm}
\includegraphics[width=1\textwidth]{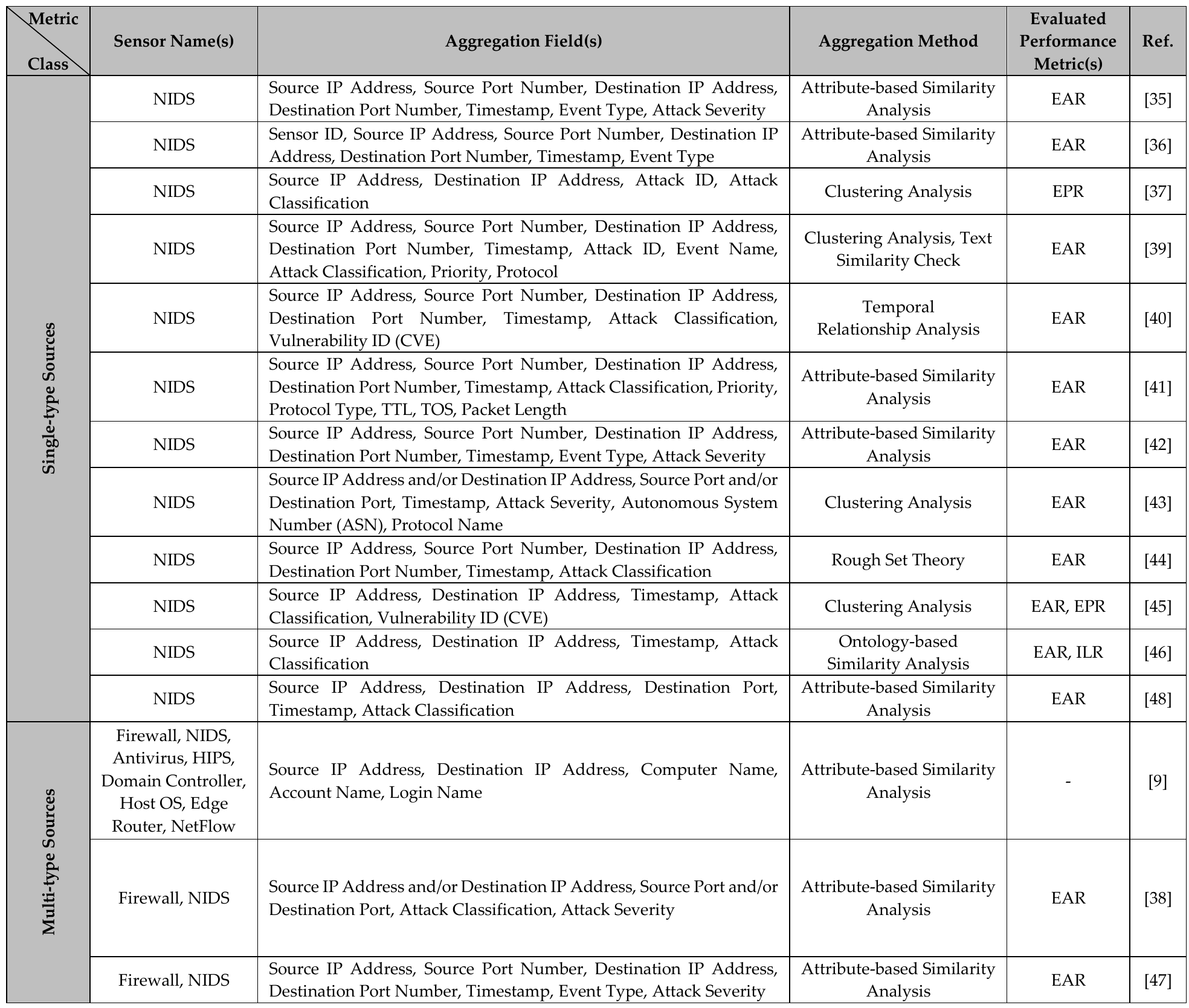}
\end{table}

Bryant and Saiedian in \cite{r9} proposed a framework to model APT attack life cycle for forensics investigation in the form of a newly introduced 7-phases IKC model that was explained in the previous sub-section. This model focuses on phases that leave trails inside the victim's network and can be observed within sensor data in different detection levels. Their proposed system reconstructs APT attack scenarios by aggregating and correlating logged low-level events by heterogeneous event log sources. During the IKC detection process, their proposed correlation component aggregates low-level events of each security sensor in a detection level regarding a perfect similarity matching process based on one or more event features. To this aim, source and destination IP addresses are used for aggregation of network-level sensors like NIDS, Firewall, and Edge Router. Also, Host/Computer Name and User/Account Name are used for aggregation of host level sensors like Host OS, HIDS, and Antivirus. In addition, for aggregating Application-level services like DNS, Email, and Web Servers, the IP address is used for similarity analysis.

Ramaki et al. in \cite{r35} and \cite{r47} and also Soleimani and Ghorbani in \cite{r42} have been proposed two alert correlation frameworks that are similar to each other to detect multi-step attacks based on multi-layer episode mining algorithms. The input of the two developed systems is IDS alerts that are produced by the Snort IDS. During the correlation process by the proposed framework, one of the main components is the aggregation component for constructing hyper-alerts by fusing low-level alerts and removing duplicated and unrelated alerts. The proposed aggregation component in these frameworks combines all of the individual alerts that have the same attack types and creates a synthetic hyper-alert for each attack type based on an aggregation time interval as the time windows. The resulted hyper-alert is an inclusive alert that means it includes all similar alerts in a specified time window. By applying attribute-based similarity analysis, they have succeeded to obtain an acceptable level of event aggregation ratio (EAR), about 99.94\% for the Snort alerts of multi-step attack scenarios in the DARPA dataset. However, their approach is limited to the NIDS sensor and does not apply to a network environment with multi-sensor, disparate and heterogeneous sensors.

Another research work is done by Husak et al. \cite{r36}. The authors in this paper proposed an alert aggregation framework for NIDS alerts. They presented four use-cases of aggregation of alerts from NIDS sensors namely, duplicated alerts, continuing alerts, aggregating alerts from overlapping sensors (overlap in their detection scope), and aggregating alerts from non-overlapping sensors. They used an attribute-based similarity analysis for the alert aggregation process based on the main IDS alert features namely, source IP and port number, destination IP and port, event type, sensor ID, and timestamp. By doing some useful case studies on a private network, they found that the system can aggregate volumes of low-level alerts up to 85\% in the EAR rate. Similar to this work, Nadeem et al. \cite{r48} proposed an attack graph generation process that has an alert aggregation component. This component is responsible for aggregating the IDS alerts, as a result of which hyper-alerts or attack episodes are generated. By using attribute-based similarity analysis on the alerts, their method yields a reduction ratio to 99.98\% of the original alert set.

Spathoulas and Katsikas \cite{r37} proposed an alert post-processing framework to improve the quality of generated alerts by the NIDS sensor using alert aggregation. The input of their developed system is alert sets of multiple NIDS sensors. Their proposed system operates in three main phases, 1) preparation phase for merging the different received alert sets by various NIDS sensors into one alert set including aggregated alert with minimum information redundancy, 2) clustering phase for creating clusters of similar events by using a set of defined similarity functions for each of event features, and 3) visualizing phase for graphically representing generated clusters to depict overall security status of the monitored network. The alert aggregation based on each attack ID in the first phase for creating aggregated alert sets and then clustering analysis on the output alert sets based on some similarity functions provides insights for the main explicit and implicit events that may be happening on the monitored network.

Fredj in \cite{r38} has been proposed a global alert correlation system that receives heterogeneous security alerts from network-level sensors i.e. NIDS and Firewall, unifies them by a customized event normalization format, discards false and unrelated alerts, aggregates remaining important once, correlates normalized generated alerts for attack scenario reconstruction. For minimizing the number of generated alerts by the security sensors and also decreasing the overall overhead of the alert correlation process, the use of an IP-based alert aggregation component that works based on a set of aggregation rules i.e. similar source IP, similar destination IP, and similar C class network address to create alert clusters. Although, it claims that the aggregation component can deal with heterogeneous log sources but they are limited only to the network level sensors.

Sun and Chen in \cite{r39} have been developed an IDS alerts aggregation system based on the frequent pattern growth-based approach. Their proposed system consists of the three main components namely, removal of noisy data, mining association rules, and text similarity check. They used the density-based spatial clustering of applications with noise (DBSCAN) algorithm for removing noise alerts. After filtering the noise data, they applied the FP-Growth algorithm to mine association rules based on a predefined time interval. By using these rules, the alerts are grouped into one meta-alert. Then, a text-similarity analysis on the resulted meta-alerts outputs the final aggregated alert set.

Alserhani in \cite{r40} has been provided an alert correlation framework for detecting multi-step attack scenarios called MARS. In the MARS engine, one of the main components is event aggregation. This component is responsible for fusing a series of NIDS alerts that are related to a single step of a multi-stage attack based on a temporal relationship. To this aim, they used a predefined window time for determining whether two generated alerts by the sensors are close enough to be aggregated into a single alert. In the MARS framework, low-level alerts are modeled into a directed acyclic graph (alerts as vertices and casual relationships between alerts as edges). Further aggregation based on a graph reduction technique removes duplication in graph vertices. Their experimental results on the DARPA dataset show that their method is capable to reduce Snort alerts by a maximum EAR rate of about \%98.6 under certain conditions.

Shittu et al. in \cite{r41} have been proposed a comprehensive system for analyzing intrusion alerts, called ACSAnIA, which operates some post-correlation analysis on generated alerts set by a set of NIDS sensors. These analyses include clustering and prioritizing alerts. One of the key benefits of the ACSAnIA framework is that it aggregates low-level alerts using similarity analysis to construct meta-alerts. They are a group of alerts that have identical values for the specified event features i.e. intrusion type, signature ID, source and destination IP addresses, and source and destination port numbers. In this paper, for evaluation purposes, a cluster quality metric, called silhouette coefficient, is used for identifying well-clustered alerts where there is a high intra-similarity between the alerts of each cluster but a low inter-similarity between the various alert clusters.

Carlisto et al. in \cite{r43} proposed multi-step attack extraction models using process mining-based approaches. Based on their proposed method, the two main concepts in the area of process mining are mapped to the area of intrusion alert analysis, activity and case (process instance). To this aim, each of the logged events by the various NIDS sensors corresponds to an activity and a case is multiple events that are grouped which share the same information. In their proposed approach, the aggregation component is responsible to generate possible cases regarding common features of the alerts. Each of the cases is related to a single step of a complex attack. There are two main aggregation strategies in this work, one-to-many (alerts with one single source IP address as attacker and many destination IP address as targets) and many-to-one (alerts with many source IP address as attackers and one single destination IP address as target).

Zhang et al. in \cite{r44} proposed an alert aggregation framework for network-level sensors (NIDS and Firewall) based on the Rough Set Theory. The proposed framework consists of the four main steps as follows: 1) Alert normalization for converting logged event formats to the standard IDMEF format to have a unified set of events, 2) Feature weight assignment for computing the importance of each event feature according to information provided by it, 3) Similarity analysis for computing similarity between two received alerts based on a predefined threshold, and 4) comparing time interval of alerts where if two detected similar alerts (similarity $\ge$ threshold) by the previous component occurred at a very close time interval, they are aggregated. Their experimental results show that their method is capable to reduce low-level alerts by a maximum EAR rate of about \%98.

Kim et al. in \cite{r45} proposed a scalable security event aggregation system over MapReduce, called SEAM-MR. Their proposed system is based on big data technologies to deal with large-scale security data which are generated during modern attacks i.e. APTs. The SEAM-MR contains three core functions namely, 1) periodic aggregation for collecting and aggregating events within the last time window, 2) on-demand aggregation for accessing users to the system for the aggregation analysis, and 3) query support for effective analysis to retrieve aggregated events for situation analysis. They presented seven use-cases of aggregation of events from NIDS sensors regarding common features of them i.e. attack source, attack target, attack source group, attack target group, and attack class. The aggregation engine provides a set of MR functions for the seven situations. They evaluated their method on a Hadoop cluster by using a variety of synthetic datasets with different properties and reported that the SEAM-MR system decreases the volume of some datasets up to 85\%.

Saad and Traore in \cite{r46} have been developed an IDS alert aggregation component to tackle the alert flooding problem. To this aim, they proposed a new alert aggregation and reduction technique based on semantic similarity between IDS alerts. The key idea in the proposed technique is that alerts that correspond to the same attack instance are semantically similar, even they have different formats. Regarding the developed approach, semantic similarity measures for every two different alerts based on the attributes of an IDS alert by proposing an ontology for the intrusion detection domain. The key benefit of this approach is information preservation during alert aggregation purposes. Also, the authors proposed a new metric, information loss ratio (ILR), and shown their system has a lower level of ILR metric compared to the traditional aggregation techniques. Although ILR is a good metric for aggregation components, their ontology is limited to the IDS alerts. In this paper, we try to expand the ontology for all event types of different sensors in all the three mentioned detection levels.

\vspace{-2mm}
\subsection{Discussion}\label{2.5.disc}
In this section, the most important results obtained from the literature review are discussed. The discussion is provided to understand the main challenges and issues of the existing works and also the requirements and capabilities of a recommended event aggregation method that is capable of reducing the volume of logged events by heterogeneous sensors. The main remarkable results are as follows:
\begin{itemize}
\item After studying the details of the existing IKC models in Section~\ref{2.1.ikc}, it can be found that the Bryant kill chain presents the common attack stages of them used in APT attacks. Hence, this IKC model is chosen for further usages related to the aggregation analysis. In addition, a provided publicly available dataset by the presenters of this model will be a useful case for the evaluation purposes of the proposed method. 

\item By investigating the layered security model and finding its relationship to the common stages of the IKC models used in APT attacks (Section~\ref{2.2.lay}), it can be found that the main detection levels as layers of security for tracking and mitigating the intrusion activities during the stages of IKC model are Network, Host, and Application. Hence, based on these detection levels, we present a set of heterogeneous security and non-security sensors that are useful for logging APT attack symptoms during the IKC stages and their related attack steps (Section~\ref{2.3.sen}). Besides, for proposing our event aggregation method, a good event feature analysis is done based on the suitable sensors related to the each of detection levels.

\item According to the presented analysis in Table~\ref{tbl:1} (Section~\ref{2.4.agg}), we found that many different techniques have been used in the aggregation component of an alert/event correlation system including statistical, clustering, and attribute-based similarity analysis. Generally, the main shortcoming of the existing machine learning-based methods is their low flexibility to adapt to the needs of users and experts for the degree of aggregation. In addition, these methods are impractical for conducting aggregation analysis with massive data generated by heterogeneous sensors due to high computational cost and poor performance. Moreover, it could be inferred that some research works suffer from producing irrelevant aggregated events due to failure to remove false-positive events that cause incorrect scenarios of attack in later stages of an event correlation system.

\item Regarding the provided comparison in Table~\ref{tbl:1}, it can be found that the related works can be classified into two main categories, namely, single-type event source and multi-type event source aggregation. Single-type event source class introduces those research works which have been proposed aggregation techniques for logged events by a unique security device \cite{r35}-\cite{r37}\cite{r39}-\cite{r46}. The main concentration of this class is the network-based IDS (NIDS) sensors. In contrast, the related works to the multi-type event source class aim to aggregate heterogeneous events which are produced by different devices of a monitored network \cite{r9}\cite{r38}\cite{r47}. These works intend to combine a series of events with similar event features which refer to attack stages related to the same activity. However, most research works have been provided a method to aggregate the events of a single-type event source i.e. NIDS and there is no suitable IKC-based aggregation method in the presence of heterogeneous sensors in the literature.

\item In the most related works only the EAR is used to evaluate the aggregation methods. However, this is a volume-centric metric and therefore could not measure the quality of output in the event aggregation methods. To the best of our knowledge, there is no suitable method in the literature considering information loss problem during the aggregation process in the presence of heterogeneous sensors. 
\end{itemize}

In this paper, to address the above-mentioned problems and issues, we propose a three-phase event aggregation method in the next section that collects, normalizes, clusters, filters, and summarizes logged events by various heterogeneous sensors during the IKC stages of APT attacks to generate aggregated events in a controllable and flexible manner.

\section{The Proposed Event Aggregation Method}
\label{3.prop}
In this section, we introduce our proposed event aggregation method. Before the description of the proposed method, it is required to explain an abstract schema of a heterogeneous event aggregation framework for understanding the position of our proposed method in the whole structure. The abstract schema is shown in Fig.~\ref{fig:4} which consists of the five main layers as follows:
\begin{itemize}
\item \textbf{Hardware (Layer 1):} in a monitored network, each of the heterogeneous sensors can be in the form of a dedicated physical machine or a virtual machine that is configured and managed by security staff for logging events.
\item \textbf{Heterogeneous Events Logging (Layer 2):} in this layer, low-level events are logged by the various sensors of the different detection levels (Network, Host, and Application) during the time. 
\item \textbf{Event Preprocessing (Layer 3):} in this layer, after the collection of logged events from heterogeneous sensors (with a variety of event log formats regarding different types and vendors of a specific sensor), they are received by an event normalization component which converts them into a common event format by using a standard normalization format which is an essential task for later processing i.e. aggregation analysis.
\item \textbf{Event Processing (Layer 4):} this layer contains a three-phase event aggregation process that receives normalized events in a specified time interval as input and after event clustering, filtering, and summarizing processes on the received events, produces final aggregated events as output.
\item \textbf{Presentation (Layer 5):} in this layer, a set of aggregated events (without any redundant information) are reported to the security administrators based on queried data to give them a complete picture of the organization's security status.
\end{itemize}

Although the main goal of this paper relates to the Event Processing Layer of the Fig.~\ref{fig:4} which is highlighted in red color, before describing the proposed event aggregation method, a brief explanation is provided on the event collection and normalization components and how they work in the abstract schema. Regarding Fig.~\ref{fig:4}, at first, low-level events logged by the heterogeneous sensors must be collected for later processing. In the field of security event collection and monitoring, various tools have been developed so far, each of which has various applications based on their functionalities and capabilities\cite{r74}. One of the most important event analyses is the ELK platform which consists of the three main parts, Elasticsearch for event gathering and indexing, Logstash for event normalization and transforming, and Kibana for event visualization and dash-boarding\cite{r51}. In the ELK architecture, there is some agent-based event collection called Beats\cite{r52}. Beats are data shippers of ELK stack which have different types for various platforms and applications i.e. Filebeat and Winlogbeat. The main reasons for choosing Beats of ELK stack to collect logged events in our method are as follows: 1) the ability to deploy agents in different detection levels (Network, Host, and Application), and 2) their high-speed capability and good performance delivering in event collection.

After collecting heterogeneous events, the gathered events may have various formats according to the variety of output log formats of a specific sensor \cite{r13}\cite{r15}\cite{r35}. Hence, before any later processing like aggregation analysis, at first, it is required to convert the event format to a unified and standard log format to understand more of the meaning of the events by the other components \cite{r13}\cite{r15}. The event normalization component is a basic component of the schema that converts all received low-level events into a common and standard log format. Till now, security researchers have been developed a normalization format for the area of intrusion detection\cite{r15}. To the best of our knowledge, one of the recently developed normalization formats which are appropriate for applying in an environment with heterogeneous sensors is object log format (OLF) \cite{r53}.

In this paper, to convert received events to the OLF format, we use the Logstash module of the ELK stack. For this purpose, based on the output of each sensor in the system, first, the set of required information fields (features) is determined based on the OLF format for normalization. Next, by defining a regular expression in the form of a well-defined Regex pattern for each type of sensor at different detection levels, important fields of that sensor event are extracted. For example, this operation for the output of Snort an IDS sensor from the network detection level is depicted in Table~\ref{tbl:2}. The used Regex pattern in this figure is a Grok pattern, a facility of Logstash module \cite{r1}. It should be noted that each sensor has its own Grok pattern for event normalization purposes. After extracting the main information fields from the logged events by a unique sensor, they can be stored in a CSV file format with columns names of OLF format related to each security sensor for further usages.

\begin{landscape}
\vspace*{1mm}
\begin{figure*}[ht]
\captionsetup{justification=centering}
\centering
  \includegraphics[width=1.4\textwidth,height=0.8\textheight]{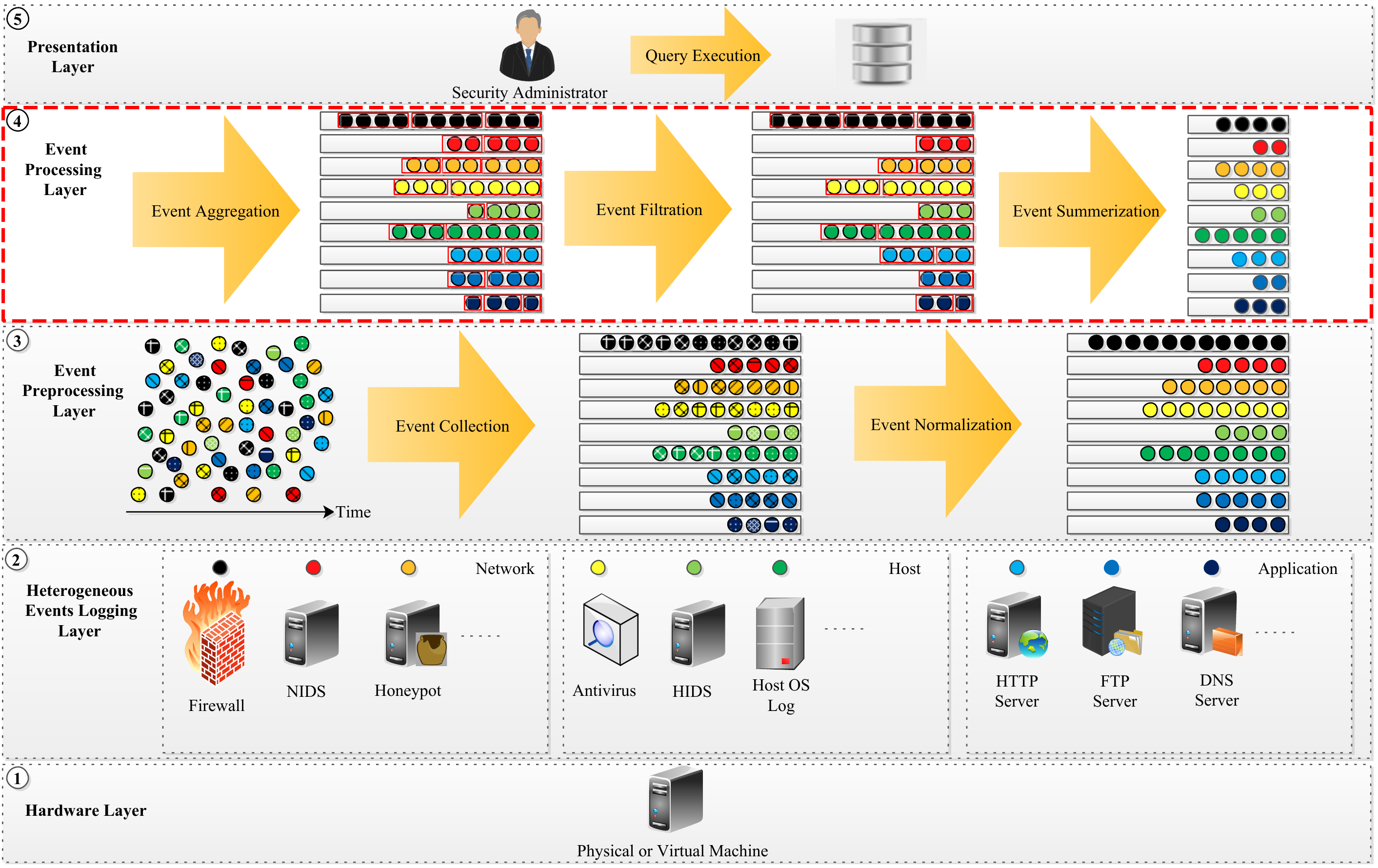}
\caption{An abstract schema of a heterogeneous event aggregation framework}
\label{fig:4}       
\end{figure*}
\end{landscape}


\begin{table}[ht]
\centering
\captionsetup{justification=centering}
\caption{Sample converted Snort IDS event type of alert to the OLF format} \label{tbl:2}
\vspace*{-4mm}
\includegraphics[width=0.8\textwidth]{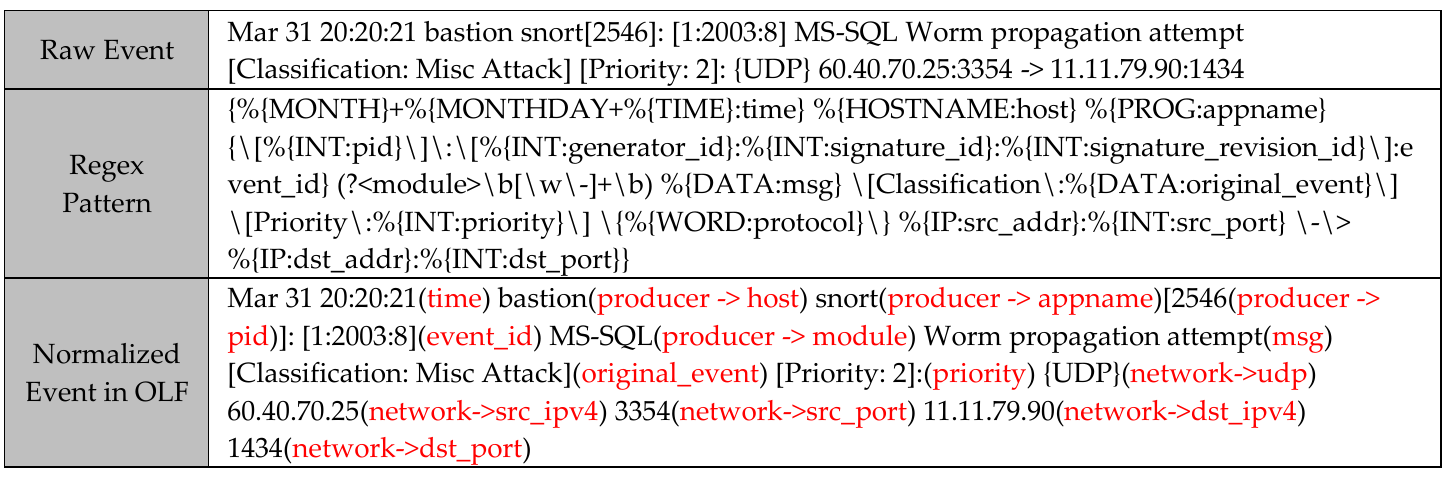}
\end{table} 

\begin{figure*}[h]
\captionsetup{justification=centering}
\centering
 \includegraphics[width=0.8\textwidth]{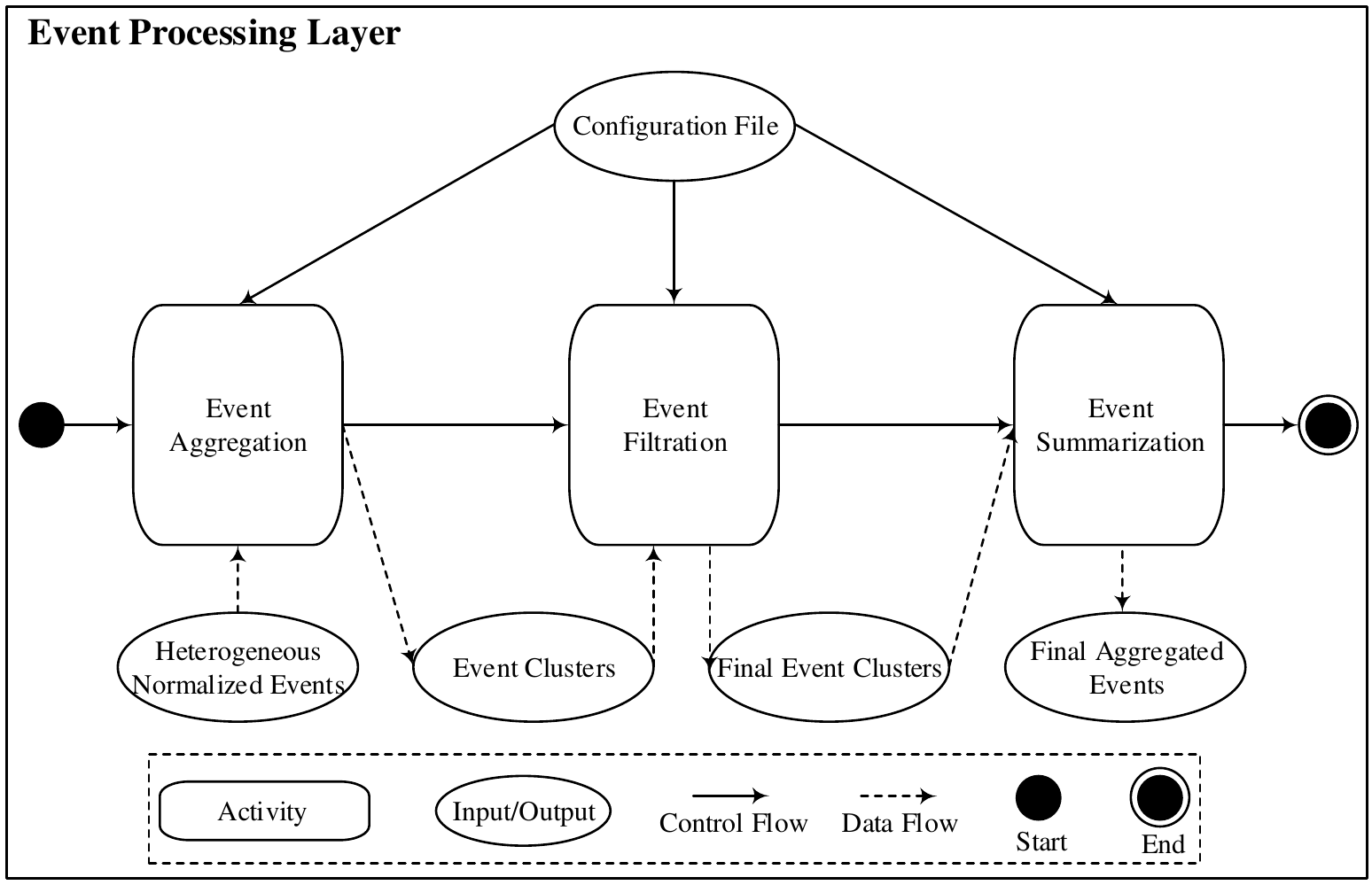}
\caption{The workflow of the main components in the proposed event aggregation method}
\label{fig:5}       
\end{figure*}

After a brief explanation of the event preprocessing layer components in the abstract schema, in the rest of this section, a detailed description of the three main components of the proposed event aggregation method namely, event aggregation, event filtration, and event summarization (event processing layer of abstract schema in Fig.~\ref{fig:4}) are provided. It should be noted that during the description of the components, as needed, the events logged in the sample attack scenario presented in the Table~\ref{tbl:4} are used for understanding the input, process, and output of the three above-mentioned components. The workflow of the main components in the event processing layer is shown in Fig.~\ref{fig:5} which are described in the following subsections.

\subsection{Event Aggregation Component}
\label{3.1.aggr}
The event aggregation component works based on the clustering analysis technique which is shown in the Algorithm~\ref{alg:1}. As illustrated in the algorithm, the algorithm receives a set of various sensor types (each sensor has a unique sensor-ID), a pre-defined aggregation time window (\textit{ATW}) for event streaming, a set of received heterogeneous events in the ATW, and a configuration file as input. The configuration file contains two types of event feature set for each sensor namely, the non-summarizable feature set \textit{NSFS} (more important features of a unique sensor used in the event aggregation component) and the summarizable feature set \textit{SFS} (less important features of a unique sensor used in the event summarization component, Section~\ref{3.3.summ}) and a time window length (\textit{TWL}) of each sensor for aggregation analysis. After the aggregation process, the algorithm generates a set of aggregated events as output and sends them to the next process. In the event aggregation component, after receiving an event set in an ATW and retrieving the parameters from the configuration file, the heterogeneous events are classified into distinct groups based on the sensor-ID. Then, the events of each group with a specified sensor-ID are sorted based on the event timestamp. Afterward, based on the \textit{NSFS} of the sensor-ID, a set of aggregation rules is created. The aggregation process is done based on the aggregation rules regarding the \textit{TWL} of the sensor-ID. The basic philosophy of this process is that if two events of a unique sensor have the same value for the \textit{NSFS} while being close in time regarding the \textit{TWL}, then they can be aggregated to create an event cluster. The candidate \textit{NSFS} and \textit{SFS} of each sensor mapped to the seven Bryant IKC stages are shown in Fig.~\ref{fig:6} in Section~\ref{6.appendix}.


\begin{algorithm}[H]
 \KwData{SS, a set of heterogeneous sensors with a unique sensor-ID \\ \hspace{7mm}ATW, aggregation time window for event streaming\\ \hspace{7mm}ES, a set of received heterogeneous events in the ATW\\ \hspace{7mm}CF, a configuration file containing various sensors features information i.e. \textit{NSFS} and \textit{SFS} for each sensor and the aggregation-related threshold\\ \hspace{11mm} values}
 \KwResult{\textit{ECS}, a set of event clusters}
\vspace{2mm}
 Get ES based on ATW size\\
 Read CF\\
 Classify heterogeneous events based on sensor-ID \textit{s} in SS\\
\For{each sensor-ID \textit{s} in SS}{
         Get TWL of the \textit{s}\\
         Get events of the \textit{s} from the \textit{ES} as \textit{s-es}\\
         Sort events in \textit{s-es} based on timestamp\\
         Get \textit{NSFS} ($f_1 - f_{j-1}$) from the \textit{CF} for the \textit{s}\\
         Get a set of aggregation rule based on \textit{NSFS} ($f_1 - f_{j-1}$) as \textit{s-aggrules}\\
         Get $\epsilon$ from the \textit{CF}\\
\For{all events in \textit{s-es}}{
	Create an \textit{EID-set} from the \textit{EIDs} in \textit{s-es} as \textit{eid-s}
	\For{each \textit{EID} in \textit{eid-s}}{
	Create an an array of \textit{events[]} containing \textit{EID}\\
	base=0\\
	 \While{base $<$ events.size}{
 	   \eIf{events[base].number $>$ 0}{
  	 	current = base + 1\\\
		\While{(current $<$ events.size) $\wedge$ (timediff (base, current) $<$ $\epsilon$)}{
			\For{each aggrule in s-aggrules}{
				eventsim[] = checksim(base, current) //If \textit{base} and \textit{current} be similar, \textit{checksim} returns 1
			}
			  \eIf{eventsim = 1}{
   				events[base].number = events[base].number + 1\\
				events[base] = events[base] + events[current]\\
				events[current].number = 0
			         }{
				   current = current +1
			}	
		}
		}{
		base = base +1
	}
	\textit{ECS} = events\\
	Return \textit{ECS}
   }
  }
 }
}
\caption{$Event\_Aggregation (S, ES, ATW, CF)$}
\label{alg:1}
\end{algorithm}

For the aggregation of each sensor event using aggregation rules, at first, they are classified into different event classes according to their event type or event id (feature 8 and 9 in ). For each event class, an array of events with a unique event type is created containing a set of ordered events that are similar in event type. In other words, one array of events is created for each event type. Then, similar events of the array are grouped within a cluster that inherits the \textit{NSFS} of the corresponding aggregation rules. To this aim, for each array, a forward scan is enough to accomplish the needed event clustering. By using two indices (base and current) each event (base event) is compared to its following events (current events) in the array. If they have similar feature values according to all aggregation rules related to a specified sensor and their time difference to the base event is lower than the \textit{TWL} of the sensor-ID, the events are grouped into a unique cluster. The value of \textit{TWL} is specified according to the speed of the event generation for each of the heterogeneous sensors. Afterward, those events in the \textit{TWL} that satisfy aggregation rules are joined to the base event. After detecting two similar events, the number of the base event is incremented by 1, the current event is added to the generated cluster, and the value of the current index is set to 0 which means the event has been analyzed. After carrying out the mentioned process for distinct event types of all heterogeneous sensors, the aggregated events in the form of different event clusters with different members are passed to the next event cluster filtration component to detect and filter noisy event clusters.

\begin{figure*}[ht]
\captionsetup{justification=centering}
\centering
  \includegraphics[width=1\textwidth]{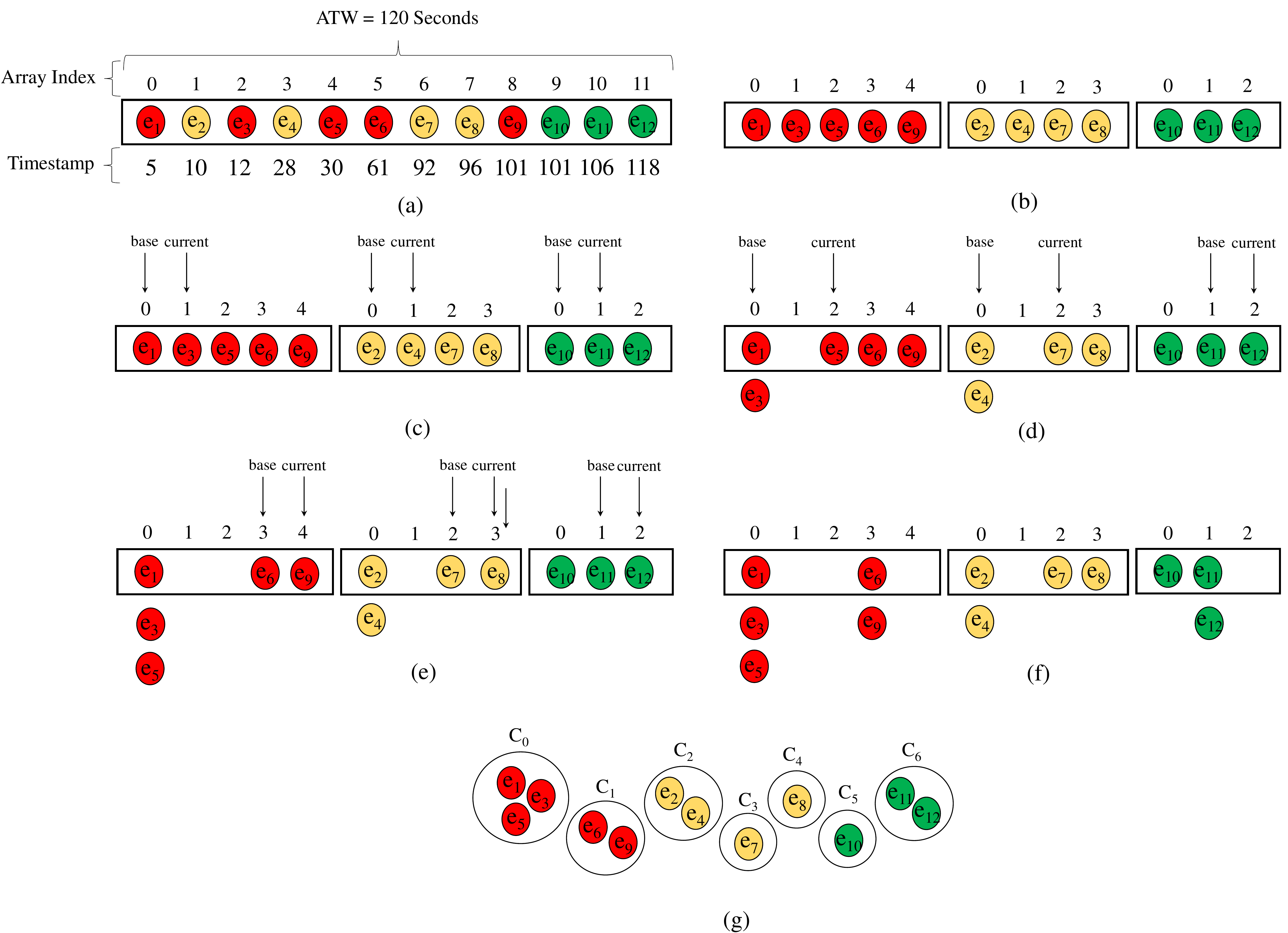}
\caption{Sample output of event aggregation component, (a) low-level event stream, (b) grouped events, (c)-(f) event aggregation (ATW=120 Sec., $TWL_{NIDS}$=$TWL_{Firewall}$=60 Sec., and $TWL_{Host OS}$=120 Sec.), and (g) resulted event clusters}
\label{fig:15}       
\end{figure*}

The output of the aggregation component for some logged events in Table~\ref{tbl:4} is illustrated in Fig~\ref{fig:15}. For the sake of simplicity, assume that the NIDS and Firewall events from the Network level ($e_1$ to $e_9$) and also the Host OS events from the Host level are considered for the aggregation analysis. Also, it is assumed that after the event collection process, the logged events, are normalized and sent to a central correlation system including the aggregation process. According to Table~\ref{tbl:4}, the NIDS, Firewall, and Host OS sensors have been logged 5 (red color), 4 (yellow color), and 3 (green color) events, respectively.

Regarding Algorithm~\ref{alg:1}, logged low-level events by the mentioned sensors are received by the event aggregation component (part (a) in Fig~\ref{fig:15}). Based on the three SIDs in Table~\ref{tbl:4}, the low-level events are categorized into three different event sets, NIDS events (IDs $e_1$, $e_3$, $e_5$, $e_6$, and $e_9$), Firewall events (IDs $e_2$, $e_4$, $e_7$, and $e_8$), and Host OS events (IDs $e_10$, $e_11$, and $e_12$) (part (b) in Fig~\ref{fig:15}). As mentioned before, in order to cluster the low-level events, some aggregation rules are used according to the \textit{NSF} set of a unique sensor type. Based on the provided information on the \textit{NSF} of the three sensors in Table~\ref{tbl:6}, the related aggregation rules are shown in Table~\ref{tbl:7} (for any two subsequent events $e_i$ and $e_j$). 

\begin{table}[ht]
\centering
\captionsetup{justification=centering}
\caption{Sample sensors features classification into \textit{NSF} and \textit{SF}} \label{tbl:6}
\vspace*{-4mm}
\includegraphics[width=1\textwidth]{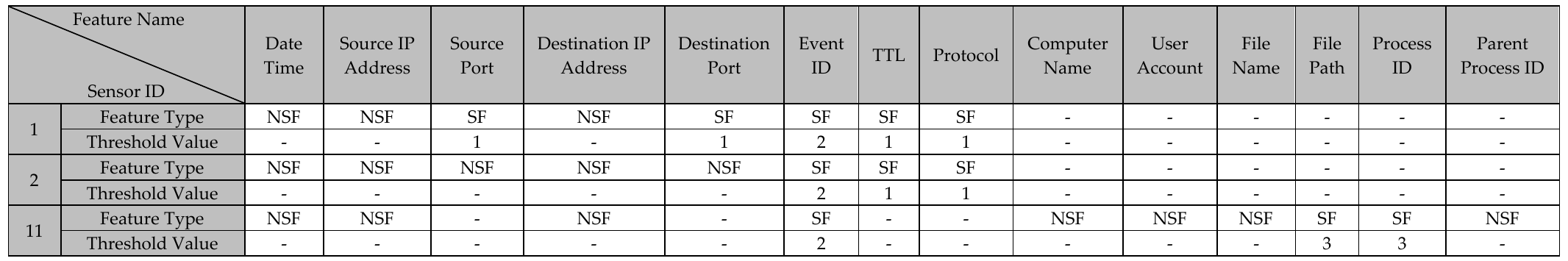}
\end{table}

\begin{table}[ht]
\centering
\captionsetup{justification=centering}
\caption{Sample aggregation rules for the NIDS, Firewall, and Host OS sensors} \label{tbl:7}
\vspace*{-4mm}
\includegraphics[width=1\textwidth]{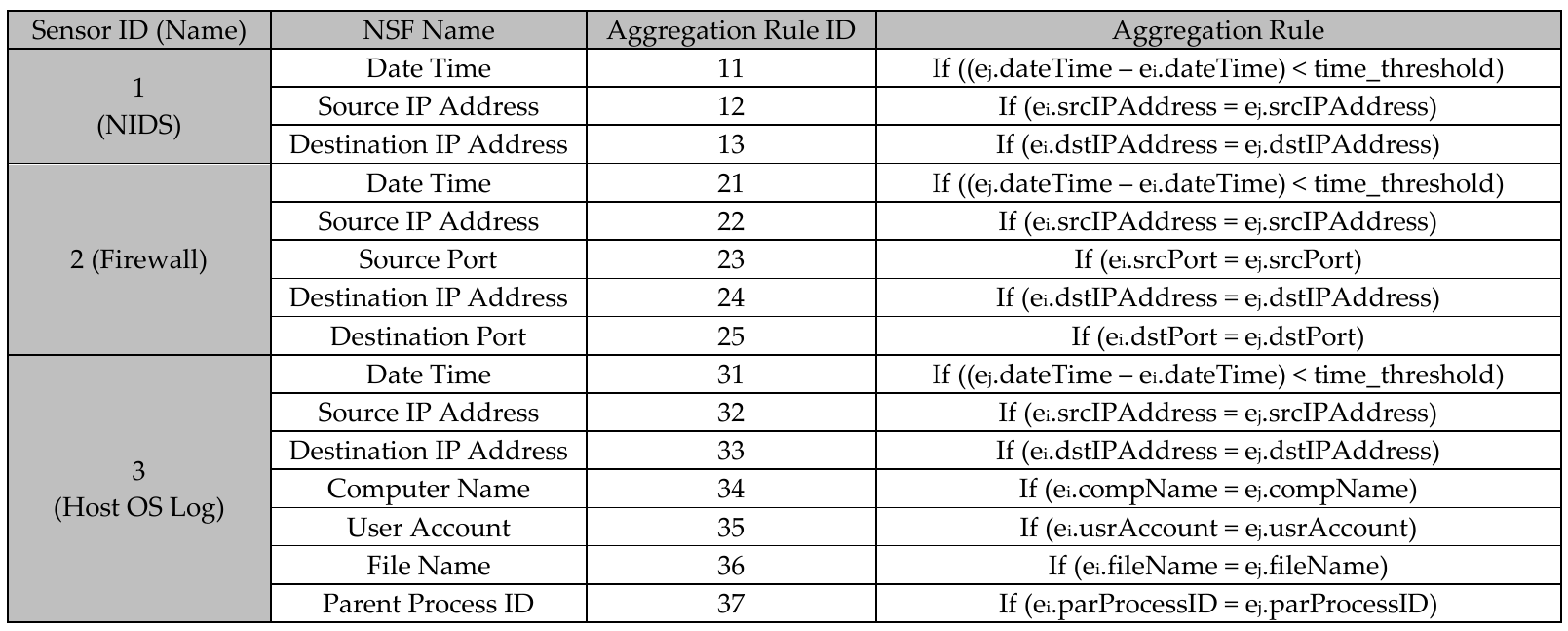}
\end{table}

So, based on the execution of the Algorithm~\ref{alg:1} on the three separated event sets, it could be inferred that the NIDS events can be clustered into two event clusters, $C_0$ ($e_1$, $e_3$, and $e_5$) and $C_1$ ($e_6$ and $e_9$), regarding aggregation rules with Rule ID 11 (with \textit{TWL}=60 seconds), 12, and 13 of Table~\ref{tbl:7}. Also, for the Firewall events, the logged low-level events are also clustered into three event clusters, $C_2$ ($e_2$ and $e_4$), $C_3$ ($e_7$), and $C_4$ ($e_8$), according to the aggregation rules with Rule IDs 21 (with \textit{TWL}=60 seconds), 22 to 25 of Table~\ref{tbl:7}. In addition, regarding Host OS events, there are two distinct event clusters, $C_5$ ($e_{10}$) and $C_6$ ($e_{11}$ and $e_{12}$), regarding aggregation rules with Rule IDs 31 (with \textit{TWL}=300 seconds), 32 to 37 of Table~\ref{tbl:7}. The event aggregation analysis and also the resulted event clusters are depicted in Fig~\ref{fig:15} from part (c) to part (g).

\subsection{Event Filtration Component}
\label{3.2.filt}
After generating event clusters from low-level events, it is time to identify significant events within a large set of logged events and filter noisy ones because they are false-positive and irrelevant events. These events are generated due to many causes like misconfiguration, low accuracy of applied detection methods, and lack of attention to contextual information of target during the event generation \cite{r15}. In the literature, there are many techniques for event verification and filtration \cite{r13}\cite{r15}. One of the most important ones is using outlier detection-based approaches \cite{r54}. In this paper, we use local density cluster-based outlier factor (LDCOF)\cite{r55}, an extension of clustering-based local outlier factor (CBLOF)\cite{r56}. Based on this metric, the system can detect noisy event clusters and filter them. A noisy event cluster is a cluster that their events are recognized as the events belonging to none of the event clusters. Detecting and filtering noisy clusters is useful to have a set of high-quality real event clusters for the next processing step. The procedure of the event cluster filtration algorithm is depicted in the Algorithm~\ref{alg:2}.

Regarding Algorithm~\ref{alg:2}, for detecting outlier event clusters, the component receives generated event clusters set (\textit{ECS}) by the previous component and the configuration file as input and generates a set of aggregated events for each sensor. After receiving \textit{ECS}, at first, they are sorted based on the number of events in each event cluster as shown in the figure. By considering the two numeric parameters $\alpha$ and $\beta$, the generated clusters are divided into two sets, large clusters (\textit{LC}) and small clusters (\textit{SC}), regarding the size of clusters. Suppose $|C_1|>|C_2|>|C_3| > ... > |C_k|$, then \textit{b} is the boundary of a cluster, small or large, such that $|C_1|$+$|C_2|$+ ... +$|C_b|$ $\ge$ $|D|$ * $\alpha$ or $|C_b|$/$|C_{b+1}|$ $\ge$ $\beta$ where \textit{D} is the whole dataset. It should be noted that $\alpha$ and $\beta$ parameters are set so that the following conditions are met, 1) most of the events in the dataset are not outliers and 2) \textit{LC} and \textit{SC} should be significantly different in size. Hence, \textit{LC}=\{$C_i | i \leq$ b\} and \textit{SC}=\{$C_j | j >$ b\}. After identifying \textit{LC} and \textit{SC} clusters, each event of a cluster is assigned with the LDCOF factor. The related equation to the this factor is shown in Equation~\ref{equ:01} where the used average distance of cluster in this equation is computed based on the Equation~\ref{equ:02}. This factor is calculated according to the size of the cluster and the distance between the target event and its nearby cluster. The rationale behind this is that when small clusters are considered outlying (noisy), the events within the \textit{SC} are assigned to the nearest \textit{LC} which becomes its local neighborhood. Thus, the events in \textit{SC} are all outliers when compared with those in \textit{LC} and discards from the \textit{ECS}. Finally, the remaining event clusters are passed to the next component as the final aggregated event set (\textit{AES}). In our example, event clusters $C_0$, $C_1$, $C_2$, and $C_6$ identified as large clusters, while $C_3$, $C_4$, and $C_5$ are considered to be small clusters. Thus, by setting $\alpha$ = 0.75 and $\beta$ = 1, it can be inferred that the events in clusters $C_3$ and $C_5$ are outliers and can be discarded.


\subsubsection{Event Summarization Component}
\label{3.3.summ}
After completion of the aggregation component, the resulted event clusters are injected into the summarization component to find duplicated events and eliminate redundant ones. The procedure of the event summarization component is shown in the Algorithm~\ref{alg:3}. Based on the figure, this component receives the sensor-ID of a sensor, a set of clustered events with the same sensor-ID, and the configuration file as input and returns a set of summarized events as output. Besides the above-mentioned information about the configuration file, it also contains concept trees of the different sensors \textit{SFS} which is used for the summarization component based on an AOI which gives extra flexibility over traditional machine learning techniques for data fusion. According to the \cite{r49}\cite{r50}, a concept tree describes the abstraction relationship (i.e. generalization/specialization) between similar concepts using a hierarchical structure. In the concept tree of an attribute, the root node contains the most abstract form of the concept and the intermediate and leaves nodes of the tree represent refined concepts and instances, respectively.

\begin{figure*}[ht]
\captionsetup{justification=centering}
\centering
\captionsetup{justification=centering}
\includegraphics[width=1\textwidth]{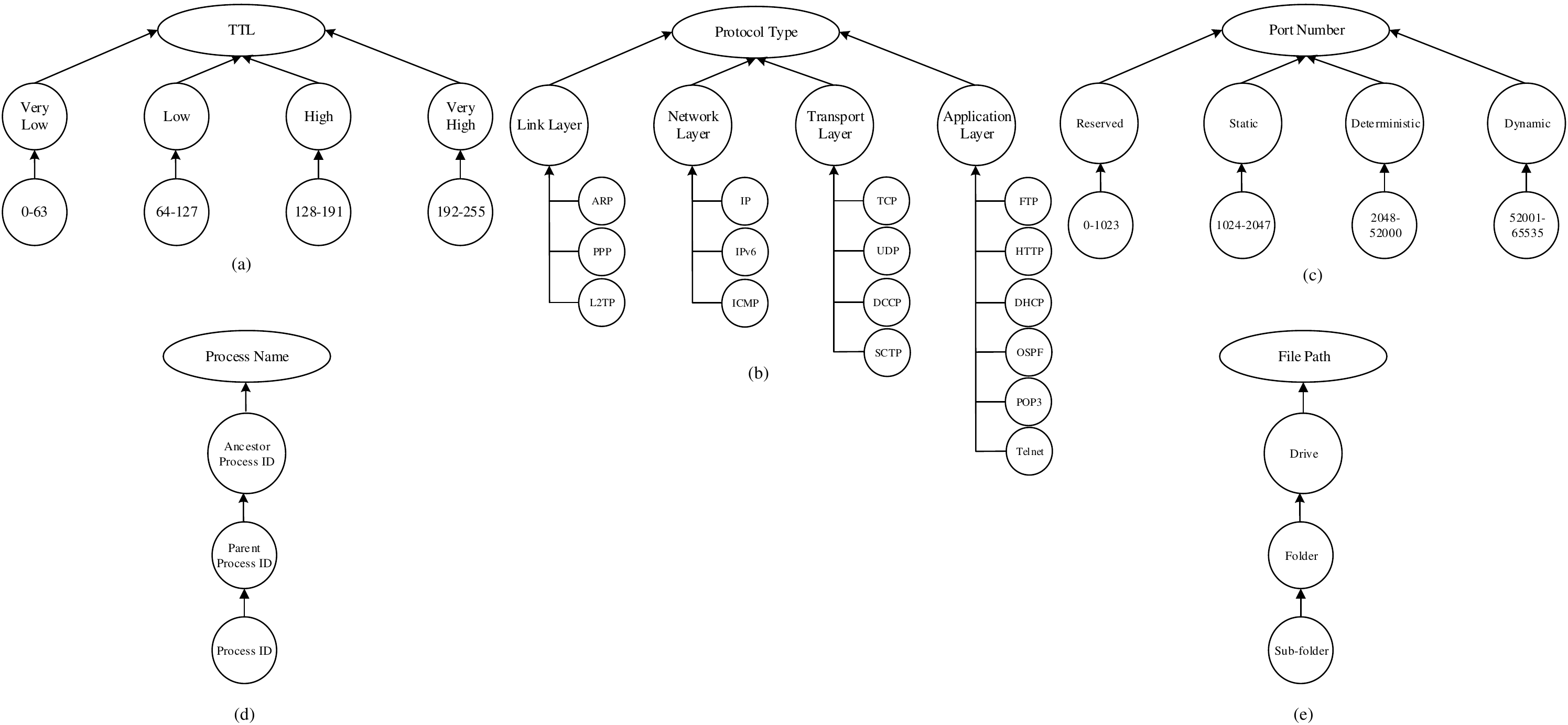}
\caption{Sample concept trees for some \textit{SF} features, (a) TTL, (b) Protocol Type, (c) Port Number, (d) Process Name, and (e) File Path}
\label{fig:8} 
\end{figure*}

\begin{figure*}[ht]
\captionsetup{justification=centering}
\centering
\includegraphics[width=1\textwidth]{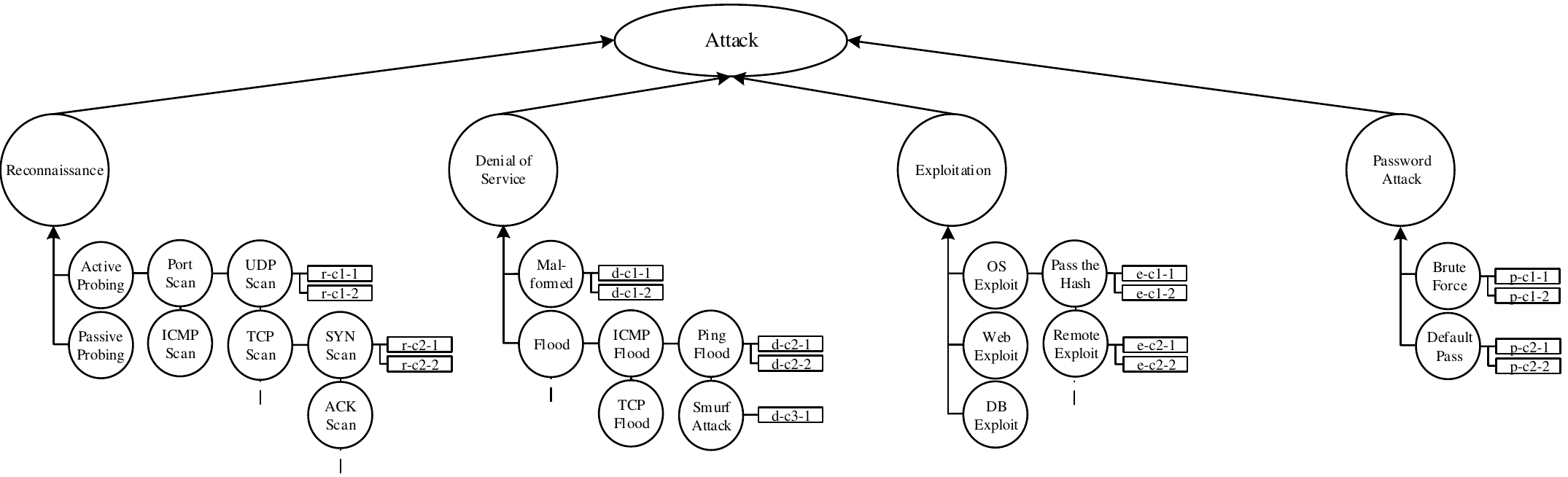}
\caption{Sample concept tree for the Event ID feature}
\label{fig:9} 
\end{figure*}

In the summarization component, for each summarizable feature (\textit{SF}), there is a unique concept tree that the feature name is the root node while the attribute values correspond to the leaves of the tree. There are some examples of the concept trees for different event features in Fig.~\ref{fig:8} and Fig.~\ref{fig:9}. As mentioned in the Algorithm~\ref{alg:3}, for a given clustered event from the aggregation component, at first, the \textit{SF} of the corresponding sensor type is obtained with the related threshold vector. The threshold vector of the sensor indicates the level of summarization process for each \textit{SF} of it according to the depth of \textit{SF} concept tree. In other words, based on the depth level of the concept tree, the event summarization method operates by the defined threshold vectors in a completely flexible manner. After retrieving the \textit{SFS} and their associated threshold vectors, for each \textit{SF}, the related concept tree of the feature is extracted from the configuration file. In the extracted tree, the relationship between children and parents is shown as a directed edge from child to parent. Then, the following tasks are repeated for each \textit{SF} until nothing else is left in the \textit{SFS} vector.

\begin{algorithm}[H]
 \KwData{\textit{ECS}, a set of clustered events with the same sensor-ID\\ \hspace{7mm}CF, a configuration file containing various sensors features information i.e. \textit{NSFS} and \textit{SFS} for each sensor and the aggregation-related threshold\\ \hspace{11mm} values}
 \KwResult{\textit{AES}, a set of aggregated events}
\vspace{2mm}
 Get \textit{ECS} //\textit{ECS} = $\{C_1, C_2, C_3, ..., C_k\}$\\
 Sort event clusters in \textit{ECS} based on their size //$\|C_1\| \geq \|C_2\| \geq \|C_3\| \geq ... \geq \|C_k\|$\\
 Get $\alpha$ and $\beta$ parameters\\
 Compute \textit{LC} and \textit{SC} based on the $\alpha$ and $\beta$\\
 \textit{AES} = $\phi$\\
\For{each event e in \textit{ECS}}{
  read current\
  \eIf{e belongs to $C_i \in \textit{SC}$}{
   Compute \textit{LDCOF} score of \textit{e} based on the Equation~\ref{equ:01}\\
   \textit{ECS} = \textit{ECS} - \textit{e}
   }{
   Compute \textit{LDCOF} score of \textit{e} based on the Equation~\ref{equ:01}\\
   \textit{AES} = \textit{AES} $\cup$ \textit{e}
  }
\textit{AES} = \textit{ECS} $\cup$ \textit{AES}
}
Return \textit{AES}
\caption{$Event\_Cluster\_Filtration (\textit{ECS}, CF)$}
\label{alg:2}
\end{algorithm}




{\large{
\begin{equation}\label{equ:01}
  LDCOF(e) = \begin{cases}
    \frac{min(d(e,C_j))}{distance_{avg}(C_j)},
    & if\hspace{1mm}e \in C_i \in \textit{SC} \hspace{1mm}where\hspace{1mm}C_j \in \text{LC}, \\
    \frac{d(e,C_i)}{distance_{avg}(C_i)},
    & if\hspace{1mm}e \in C_i \in \text{LC}
  \end{cases}
\end{equation}

\begin{equation}
\label{equ:02}
distance_{avg}(C) = \frac{\sum_{c \in C}d(i,C)}{|C|}
\end{equation}
}}

For the examined \textit{SF}, at first, the value of the \textit{total-distinct-values} is calculated based on the different values for the \textit{SF} in the generated event cluster. Then, the value is compared with the related threshold value in the threshold vector. As long as the \textit{total-distinct-values} for a given \textit{SF} is bigger than a predefined threshold value, the \textit{SF} value of the events are replaced with the value of its parent for all events of the event cluster, otherwise, it is kept. At the end of each repetition of the generalization loop, the \textit{total-distinct-values} of the \textit{SF} is calculated again for the later comparison. In addition, if there are some duplicated events in the resulted generalized event set, similar events are fused into a unique summarized event and duplicated ones are deleted. It should be noted that if the value of \textit{total-distinct-values} is still greater than the defined threshold value, the generalization process continues. This will continue until all features in the \textit{SFS} of the sensor have been checked and no other features are left. The resulted non-repetitive events of the final round are reported as the output of the summarization component (final aggregated events).


\begin{algorithm}[H]
 \KwData{S, a sensor with a unique sensor-ID \\ \hspace{7mm}\textit{AES}, a set of aggregated events with the same sensor-ID\\ \hspace{7mm}CF, a configuration file containing various sensors features information i.e. \textit{NSFS} and \textit{SFS} for each\\ \hspace{11mm} sensor and the aggregation-related threshold values}
 \KwResult{SES, a set of summarized events}
 Get \textit{SFS} ($f_i - f_n$) of \textit{s} from the \textit{CF}\\
\For{each feature $f_j$ in SFS ($f_i - f_n$)}{
	Get \textit{$threshold\_numeric\_value$} for $f_j$\\
           Get Feature $f_j$ hierarchy from \textit{CF}
           \textit{$total\_distinct\_values$} = the number of distinct values of \textit{AES} with $f_j$ + \textit{$threshold\_numeric\_value$} for $f_j$
	 \While{$total\_distinct\_values$ $>$ $threshold\_numeric\_value\_f_j$}{
  	 \For{all \textit{ae} in \textit{AES}}{
	  \eIf{$f_j$ value of \textit{ae} has a parent in $f_j$ tree}{
	   	$f_j$ value of \textit{ae} = parent value
	   }{
	   	$f_j$ value of \textit{ae} = $f_j$ value of \textit{ae}
	  }
         }
	\textit{$total\_distinct\_values$} = the number of distinct values of \textit{AES} with $f_j$\\
		  \eIf{duplicate ae in \textit{AES}}{
	   	  merge identical events into unique \textit{ae} in \textit{AES}
		 SES = SES $\cup$ \textit{AES} 
	   }{
	   	break
	  }
     }
Return \textit{SES}
}
\caption{$Event\_Summarization (S, AES, CF)$}
\label{alg:3}
\end{algorithm}

\begin{figure*}[ht]
\captionsetup{justification=centering}
\centering
\includegraphics[width=1\textwidth]{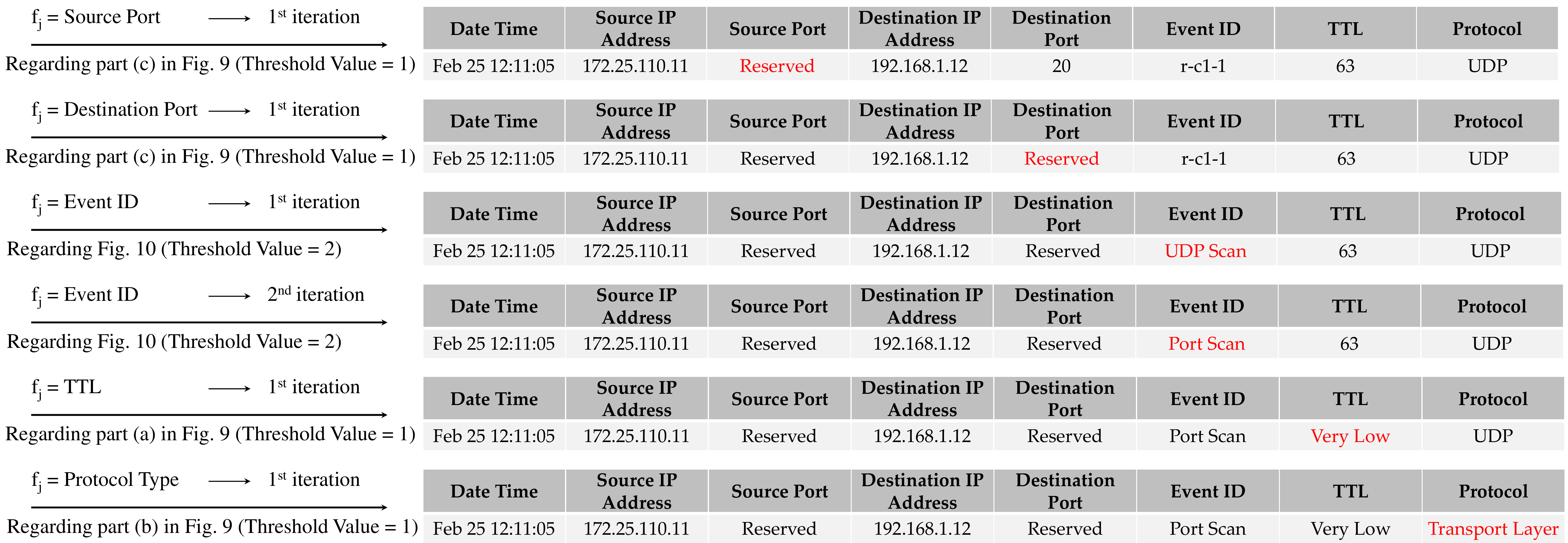}
\caption{Operation of the Algorithm~\ref{alg:3} on the NIDS sensor events in our example}
\label{fig:17} 
\end{figure*}

Regarding our example, after event aggregation and outlier event filtration processes, 5 event clusters including 10 distinct events are injected to the event summarization component. As mentioned in the Algorithm~\ref{alg:3}, for event summarization process, it is required to define the concept tree of the features in the \textit{SF} set of a specified sensor. As seen in Table~\ref{tbl:6}, the \textit{SF} set of the NIDS and firewall sensors are \{Source Port, Destination Port, Event ID, TTL, Protocol\} with threshold vector \{1, 1, 2, 1, 1\} and \{Event ID, TTL, Protocol\} with threshold vector \{2, 1, 1\}, respectively. Also, the \textit{SF} set of the Host OS is \{Event ID, File Path, Process ID\} with threshold vector \{2, 3, 3\}. Based on the defined \textit{SF} sets for the sensors in Table~\ref{tbl:6}, Fig.~\ref{fig:8} and Fig.~\ref{fig:9} indicate their related concept trees. An example of the operation of the Algorithm~\ref{alg:3} on event $e_1$ of the NIDS sensor is shown in Fig.~\ref{fig:17}.

Table~\ref{tbl:8} shows the final results of the event summarization component on the event clusters $C_0$, $C_1$, $C_2$, $C_4$, and $C_6$. According to the results of this component the pairs of events ($e_1$, $e_3$), ($e_2$, $e_4$), ($e_6$, $e_9$), and ($e_{11}$, $e_{12}$) are summarized and formed the aggregated events $ae_1$, $ae_2$, $ae_4$, and $ae_6$. In addition, the two events $e_5$ and $e_8$ are not summarized with other events and remained in the output aggregated event set. For example, the analyzes performed by the event summarization component on the NIDS sensor event clusters, $C_0$ and $C_1$, are given below.

\begin{itemize}
\item The ($e_1$, $e_3$, and $e_5$) event cluster of the NIDS: regarding the four \textit{SF} of the NIDS sensor and their related concept tree, the events of this cluster are summarized. For this purpose, at first, the Source Port and Destination Port values of the events are examined based on the Port Number concept tree (part (c) in Fig.~\ref{fig:9}). In this step, regarding the Source Port values of the three events and their parent name in the related concept tree, the value of this feature for events $e_1$, $e_3$, and $e_5$  is changed to \textit{Reserved}, \textit{Reserved}, and \textit{Deterministic}, respectively. Since, the threshold value of the Port Number is 1 (see Table~\ref{tbl:6}) and there is no parent in the concept tree for further analysis, the summarization process is finished for the Port Number feature. Then, this analysis is done for the remained \textit{SF} according to the concept trees of each of them. The results of the summarization process show that $e_1$ and $e_3$ are duplicated and so on, one of them can be eliminated, but $e_5$ remains in the final event set.
\item The ($e_6$ and $e_9$) event cluster of the NIDS: based on the aforementioned analysis for the previous event cluster and regarding the Source Port values of the two events and their parent name in the Port Number concept tree, the event values are changed to the \textit{Deterministic} value. Since, the threshold value of the Port Number is 1 (see Table~\ref{tbl:6}) and the identical values, Deterministic, for the feature Source Port, the summarization process is finished for the Port Number feature. After summarization process for the next \textit{SF} features, the results show that the Destination Port, Event ID, TTL, and Protocol are equaled to \textit{Reserved}, \textit{TCP Scan}, \textit{Very High}, and \textit{Transport Layer}, respectively for the two events. Therefore, based on the summarized events, it could be inferred that $e_6$ and $e_9$ are redundant and so on, one of them can be deleted.
\end{itemize}

\begin{table}[ht]
\centering
\captionsetup{justification=centering}
\caption{The resulted aggregated events in our example} \label{tbl:8}
\vspace*{-4mm}
\includegraphics[width=1\textwidth]{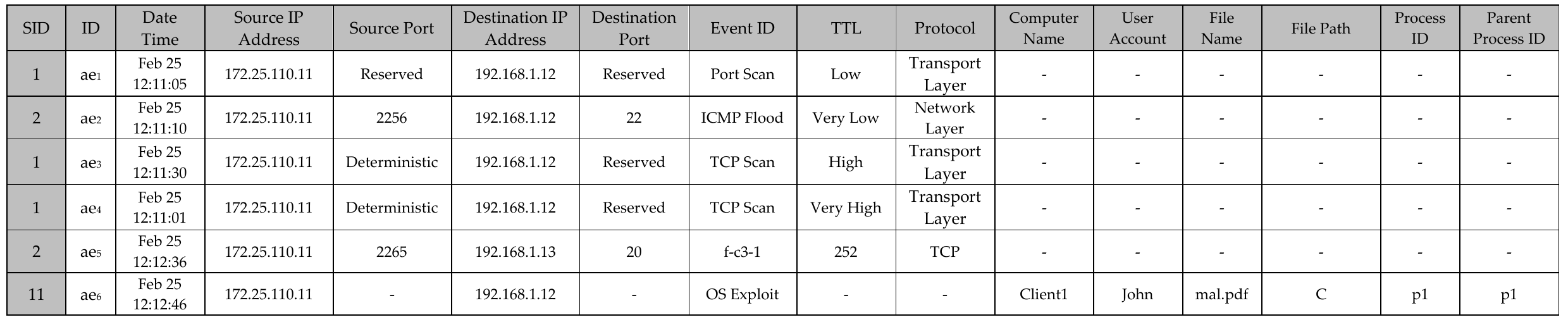}
\end{table}

\section{Evaluation and Discussion}\label{4:eval}
After explaining the proposed event aggregation method in the previous section, the evaluation method and the results of experiments are described in this section. In the rest of this section, at first, the experimental results of the simulation and numerical analysis on some standard datasets are reported in Section~\ref{4.1.simu}. In addition, Section~\ref{4.2.anal} provides some brief analytical discussion about the proposed method and its effects on the area of complex IKC-based APT attack scenarios detection and its limitations.

\subsection{Simulation and Numerical Experiments}
\label{4.1.simu}
To test the functionality and the performance of the proposed method using simulation and numerical analysis, at first, by using a comprehensive examination of the existing datasets in the area of the intrusion detection, a comparative analysis on them is provided in Table~\ref{tbl:9}, in Section~\ref{6.appendix}, which provides some useful information for each of them. This study helps us to choose the most suitable for evaluation purposes. Hence, regarding the provided information in Table~\ref{tbl:9}, the three more relevant and publicly available datasets for the evaluation are SotM34 \cite{r64}, Bryant \cite{r9}, and LANL \cite{r65}. It should be noted that all the three mentioned datasets contain a set of heterogeneous events which is logged by different sensors of the three mentioned detection levels (Network, Host, and Application).

\subsubsection{Datasets}
\label{4.1.1.data}
According to the three chosen datasets for the simulation and numerical evaluation of the proposed event aggregation method, they are explained in brief in the following from different aspects i.e. included heterogeneous sensors and event size.
\begin{itemize}
\item \textbf{SotM34:} In the area of intrusion detection, one of the main multi-source datasets is SotM34 \cite{r64} which creates by the netForensics Honeynet team during the Honeynet Project, Scan of the Month challenge. The dataset contains various heterogeneous events from three different security sensors, namely, NIDS (Snort), Firewall (IPTables), Host OS (Linux Syslog), and the only non-security sensor, Web Server (Apache). The events are logged by different sensors in a honeypot system. According to the SotM34 simulation environment, three main systems are used for recording the Honeynet activities, namely, bridge, bastion, and combo for about 4 weeks (from 22 February 2005 to 17 March 2005). The first system is used for filtering malicious connections and the second runs a NIDS service based on the well-known Snort tool. The third system is also used for the victim which contains multiple virtual IP addresses on a specified network range. In our evaluation, the low-level events logged by the different sensors (260337 individual events) are used for the experiments.
\item \textbf{Bryant:} As mentioned before, one of the main IKC models in the area of IKC-based multi-step attacks is the Bryant kill chain model \cite{r9} which is described in detail in Section~\ref{2.1.ikc}. Regarding this model, the inventors of this IKC model, Bryant and Saiedian have been released a new heterogeneous events dataset based on their proposed IKC model which is accessible upon request. In the Bryant dataset, there are 5962 distinct attack-related events for a sample multi-step attack scenario which are logged during the seven stages of the Bryant IKC model. The detailed explanation of the attack scenario and the related simulation environment (with different security and non-security sensors) is described in \cite{r9}. According to their network topology, all the events are logged by some heterogeneous sensors, namely, NIDS, Firewall, and Edge Router from the Network level, HIPS, Antivirus, Vulnerability Scanner, and Host OS from the Host level, and Domain Controller from the Application level.
\item \textbf{LANL:} One of the other main datasets containing multi-source cybersecurity events is released by Los Alamos National Laboratory that called LANL dataset \ to cite{r65}. This high volume dataset (about~10.7 GB) is publicly available, which includes events of five distinct sensors namely, NetFlow, Host OS, Audit Logs, and DNS Server with 1,648,275,307 distinct events. This dataset contains the normal behaviors of 12425 users and 117684 computers of the LANL company monitored network for 58 consecutive days. Besides logging the normal activities of the users, the dataset has a set of events for a simulated attack by a Red Team. The event log files with the related event features are described in the \cite{r65}. It should be noted that in our experiments, except for the DNS events, only part of the other sensor events (a quarter of the total) is used for the evaluation due to the large volume of events and the computational resource constraints.
\end{itemize}

\subsubsection{Evaluation Metrics}
\label{4.1.2.metr}
Besides the functional tests of the method, the other aspect of evaluation is related to the performance tests. According to the literature review, there are three main evaluation metrics for aggregation analysis which are as follows:

\begin{itemize}
\item \textbf{Event aggregation ratio (EAR):} this metric shows the power of an event aggregation method in reducing the volume of logged events by the sensors. Based on Equation~\ref{equ:1}, this metric is calculated by dividing the number of aggregated events by the number of total events in a specified ATW. The greater the value of this parameter, the greater the event aggregation rate and more event reduction. For example, in our provided example, the EAR is 50\% (total events = 12 (Table~\ref{tbl:4}) and aggregated events = 6 (Table~\ref{tbl:8})).
\begin{equation}
\label{equ:1}
EAR = \bigg(1-\frac{\# \ of \ Aggregated \ Events}{\# \ of \ Total \ Events}\bigg)*100
\end{equation}
\item \textbf{Event processing ratio (EPR):} The EPR metric indicates the processing speed of an aggregation method. Based on Equation~\ref{equ:2}, this metric is obtained by dividing the number of processed events by the processing time (in seconds). The greater the value of this parameter, the greater the processing power of the aggregation component. For example, if the proposed method process 12 events of our example in 4 seconds, the EPR equals 3 events/second.
\begin{equation}
\label{equ:2}
EAR = \bigg(1-\frac{\# \ of \ Processed\ Events}{Processing \ Time \ (in \ Seconds)}\bigg)
\end{equation}
\item \textbf{Information loss ratio (ILR):} Although in most related works only the EPR is used to evaluate the aggregation process, this is a volume metric and is not sufficient for quality evaluation of the aggregation method. The basic idea of the ILR metric is to measure the amount of security-relevant data loss during the aggregation and summarization process. As mentioned, in the proposed aggregation and summarization method, two concepts that belong to the same concept tree are summarized by replacing them with their least common ancestor (LCA) from the corresponding concept tree. This causes to occur a loss of security information. In this paper, to measure information loss by replacing the parent concept with its child concept, Equation~\ref{equ:3} is used which is borrowed from the \cite{r46}, the nearest work to our method. The ILR is a value in \[0, 1\] where 0 means no information loss and 1 indicated 100\% information loss.
\begin{equation}
\label{equ:3}
ILR(C) = \frac{\sum_{c \in C} \big(IC(c)-IC(LAC(C))\big)}{\sum_{c \in C} IC(c)}
\end{equation}
Based on this equation, parameter C is a set of given concepts. Also, for each of the concepts in an event type (indicated the type of sensor), the information content (IC) of concept c is calculated by using Equation~\ref{equ:4} which is used to measure the amount of information serves by the c concept.
\begin{equation}
\label{equ:4}
IC(c) = - log\bigg(\frac{\frac{|leaves(c)|}{|subsumers(c)|}+1}{maxleaves+1}\bigg)
\end{equation}
The value of this metric depends directly on the threshold vector values of the \textit{SFS} features set. As an example, assume that we want to compute the ILR rate coming from aggregating the two sub-classes of the ICMP Flood, namely, Ping Flood and Smurf Attack (Fig.~\ref{fig:9}). According to the Equation , we have the following IC values, IC(Ping Flood) = 1.08, IC(Smurf Attack) = 1.16, and IC(ICMP Flood) = 1.03. As illustrated in Fig.~\ref{fig:9}, since the LCA of the Ping Flood and Smurf Attack is ICMP Flood, the ILR rate is equal to ILR(Ping Flood, Smurf Attack) = 0.08 based on Equation~\ref{equ:3}.
\end{itemize}

Since, in our proposed aggregation method, clustering analysis based on the \textit{NSFS} of events plays a key role in the method, it is required to measure the quality of event clusters to find how similar the events are within a cluster and how dissimilar with the events in the other event clusters. There are many internal cluster validity metrics are presented in the area of cluster quality (CQ) to evaluate the event clustering task \cite{r66}\cite{r67}. In this paper, two related metrics are used for evaluating clustering analysis of the event aggregation method which are as follows:

\begin{itemize}
\item \textbf{Dunn index (DI):} this metric is an internal cluster validity metric which defines as a ratio of the minimum distance between clusters to the maximum cluster diameter \cite{r66}\cite{r67}. This metric aims to determine the dense and well-separated cluster which is designed for non-overlapping event clusters. This metric for a specific number of clusters is calculated using Equation~\ref{equ:5}. The details of this equation are provided in \cite{r66}. According to this metric, the larger the DI value (using Euclidian distance), the better the clustering.
\begin{equation}
\label{equ:5}
DI_{n_c} = min_{i=1, ..., n_c} \left\{ min_{j=i+1, ..., n_c} \bigg(\frac{d(c_i,c_j)}{max_{k=1, ...,n_c}diam (c_k)}\bigg)\right\}
\end{equation}
\item \textbf{Davies–Bouldin index (DBI):} this is another internal cluster validity metric that aims to find clustering algorithms with minimal intra-cluster variances and maximal inter-cluster distances. This metric for a specific number of clusters is calculated using Equation~\ref{equ:6}. The details of this equation are provided in \cite{r67}. According to this metric, the smaller the DBI value (using Euclidian distance), the better the clustering. In other words, the clusters should have the minimum possible similarity to each other.
\begin{equation}
\label{equ:6}
DBI_{n_c} = \frac{1}{n_c} \displaystyle\sum_{i=1}^{n_c} max_{j \ne i}(DBI_{ij})
\end{equation}
\end{itemize}

\subsubsection{Experimental Setup}
\label{4.1.3.setu}
By using the EPR evaluation metric, the performance of the proposed aggregation method has also been measured. In other words, the performance of the system is in terms of execution times. For this purpose, it is required to presents the experimental setup of our evaluations. The proposed method has been developed in Java and the experiments have been executed on a Windows10-based machine with the Intel Core i7 CPU (4 MB Cache and 3.2 GHz), 16 GB of RAM, and 2 TB Hard Disk.

\subsubsection{Experimental Results on the Datasets}
\label{4.1.3.expr}
\begin{figure*}[ht]
\captionsetup{justification=centering}
\centering
\vspace*{-4mm}
  \includegraphics[width=1\textwidth]{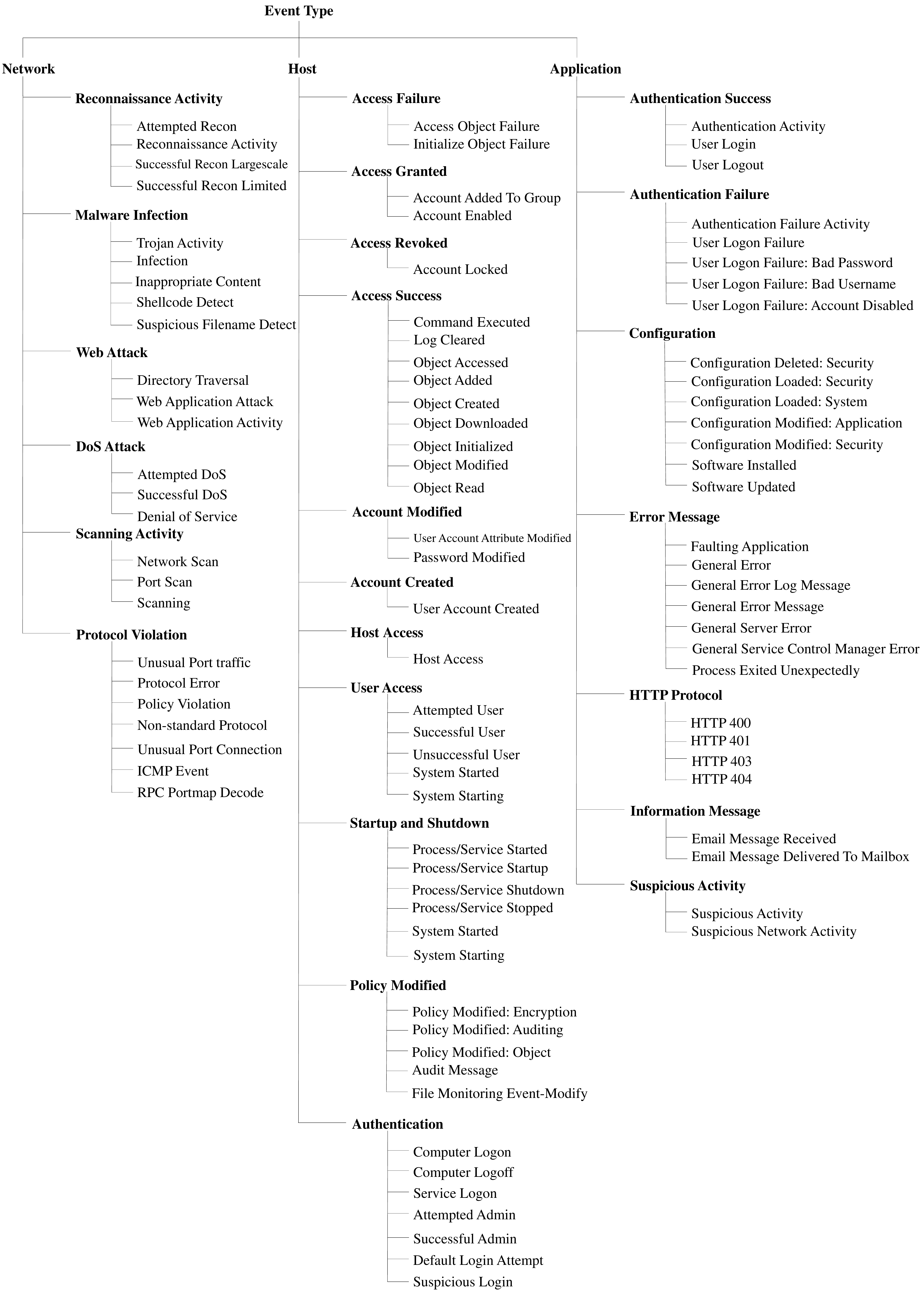}
\caption{The used concept tree for the event type feature in our experiments}
\label{fig:18}       
\end{figure*}

In this section, we present the experimental evaluation of the proposed event aggregation method according to the three chosen datasets, namely, SotM34, Bryant, and LANL. Regarding our aggregation method, there is a need for \textit{SFS} features set threshold vectors of the used sensors in each of the datasets for the event summarization component. Depending on the concept trees related to the \textit{SFS} features of each sensor in the dataset, several vectors can be defined for the threshold values. In our experiments, different threshold vectors were manually tuned and used for various sensors, the most suitable of which was selected. Table~\ref{tbl:10} in Section~\ref{6.appendix} shows the threshold vectors of various sensors of each dataset for the experiments. The \textit{SF} Numbers in the last column of Table~\ref{tbl:10}) originates from Table~\ref{tbl:3}. The used concept tree for the event type feature in the experiments is depicted in Fig.~\ref{fig:18} in Section~\ref{6.appendix}.

According to the provided information in Table~\ref{tbl:10}, the evaluation results for each of the three mentioned datasets are provided in Table~\ref{tbl:11}. As can be seen in the table, for each sensor of the datasets, a specified value is determined for the time window length (\textit{TWL}) regarding the suggested time intervals in \cite{r68} to focus on the sensors for detecting and preventing malicious activities of the attackers. After injection of the low-level events in the datasets to the proposed method with a length of 3600 seconds for the ATW parameter, some statistics related to each component of the method are reported in Table~\ref{tbl:11}. In addition, the results of the aforementioned performance evaluation metrics are also provided in this table. In addition, for filtering noisy event clusters of SotM34, Bryant, and LANL datasets in the experiments, the $\alpha$ parameter is set to 90\%, 90\%, and 80\%, respectively, an optimal value regarding each of the datasets.

Generally, regarding the various EAR and their related ILR values, the experiments have proved that the proposed aggregation method is a general solution that reduces a huge event set with a lot of redundant events into a set of summarized events with the least amount of security information loss. For example, the aggregation method has reduced a significant amount of events for some Network level sensors such as NIDS, Firewall, or Netflow, to nearly 99\% in some circumstances with a reasonable level of ILR. 

\begin{table}[ht]
\centering
\captionsetup{justification=centering}
\caption{The results of the experiments on the three standard datasets} \label{tbl:11}
\vspace*{-4mm}
\includegraphics[width=1\textwidth]{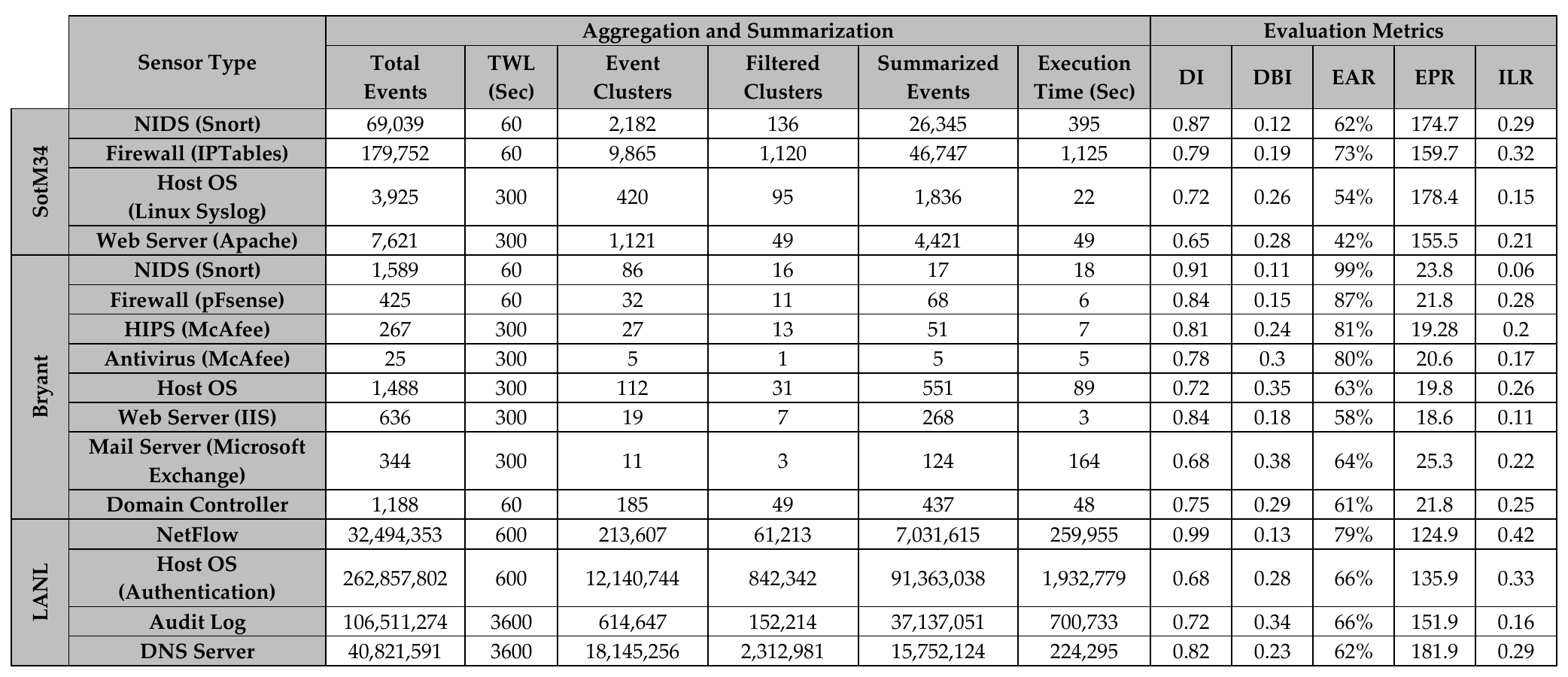}
\end{table}

Furthermore, a comparison analysis between the results of the proposed method and the related works (Section~\ref{2.4.agg}) is provided in Table~\ref{tbl:12}. Regarding the results in the table, it can be found that:
\begin{itemize}
\item The obtained EAR for the events of various sensors is very close to the values reported in the literature. This indicates that the combination of aggregation and summarization operations has been very effective in reducing the volume of logged events.
\item In terms of the number of event features used in \cite{r37}, our method has a relatively better rate for the EPR. However, in our method, in addition to the event aggregation and summarization time, the time for event normalization and filtration of noise events are also included in the reported execution time. It is worth noting that the high performance reported for the SEAS-MR method \cite{r45} is because this method is implemented and evaluated on a Hadoop cluster, which is beyond the scope of this paper.
\item The proposed method can aggregate events with minimal loss of security information, resulting in a small number of high-quality cluster events as output, that are of rich use to networks' administrators.
\end{itemize}

\begin{table}[ht]
\centering
\captionsetup{justification=centering}
\caption{Comparison performance of our method and the related works to the event aggregation} \label{tbl:12}
\vspace*{-4mm}
\includegraphics[width=0.8\textwidth]{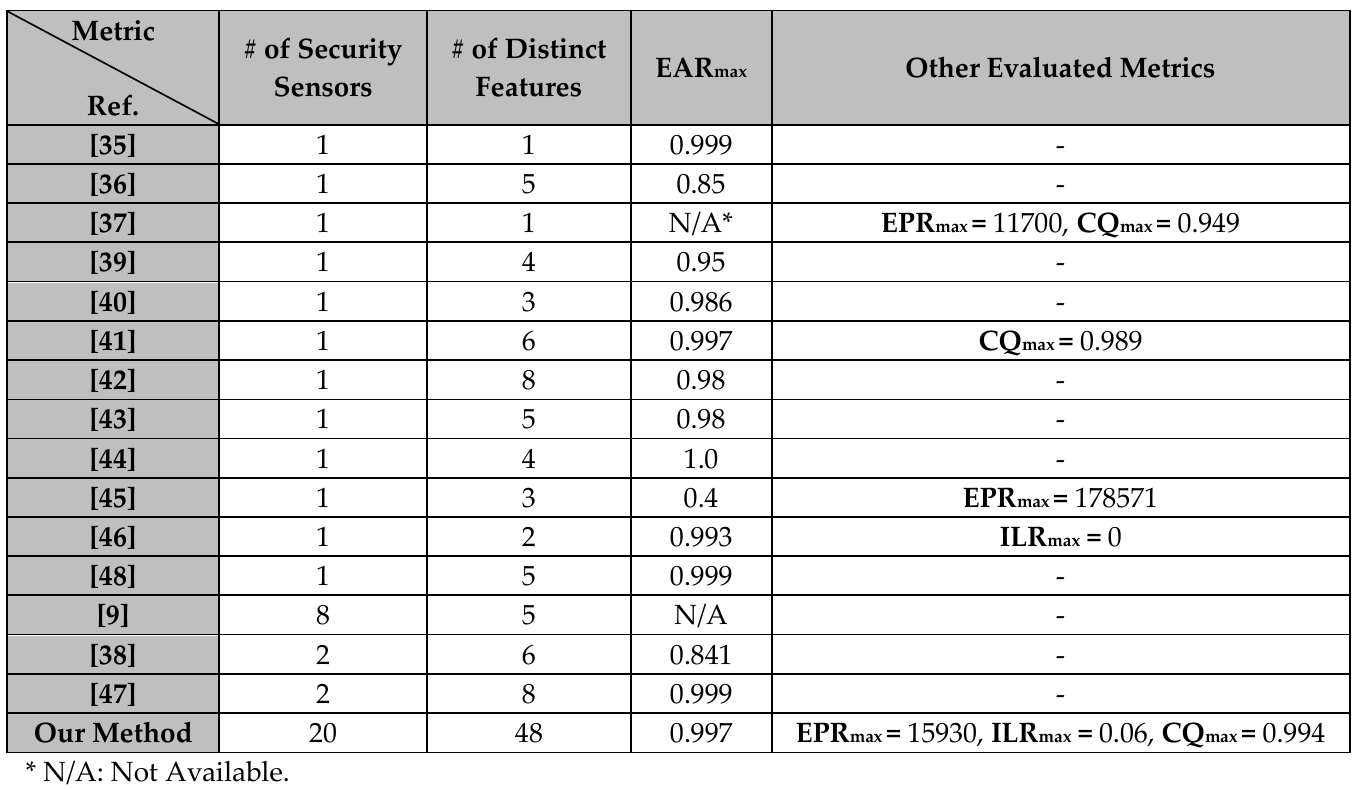}
\end{table}

Based on the results in Table~\ref{tbl:11}, it can be inferred that the performance of the proposed aggregation method can be influenced by several factors such as event generation rate by the sensor, the \textit{TWL}, the number of event features, the number of \textit{NSFS} and \textit{SFS} of a sensor, and also the threshold values for each of the system parameters i.e. \textit{SF} threshold vectors. The effects of some mentioned factors are analyzed and discussed in the next sub-section.

\subsection{Analysis and Discussion}
\label{4.2.anal}
After evaluating the proposed event aggregation method, in this section, we examine the results of several analytical reviews on the datasets and the relationship of different parameters on each other. Based on these analyses, we show how the proposed aggregation method helps to develop event preprocessing analysis such as event aggregation to design a SIEM or SOC solution.

\subsubsection{The Relationship between EAR and TWL}
\label{4.2.1.ear-twl}
\begin{figure*}[ht]
\captionsetup{justification=centering}
\centering
  \includegraphics[width=0.7\textwidth]{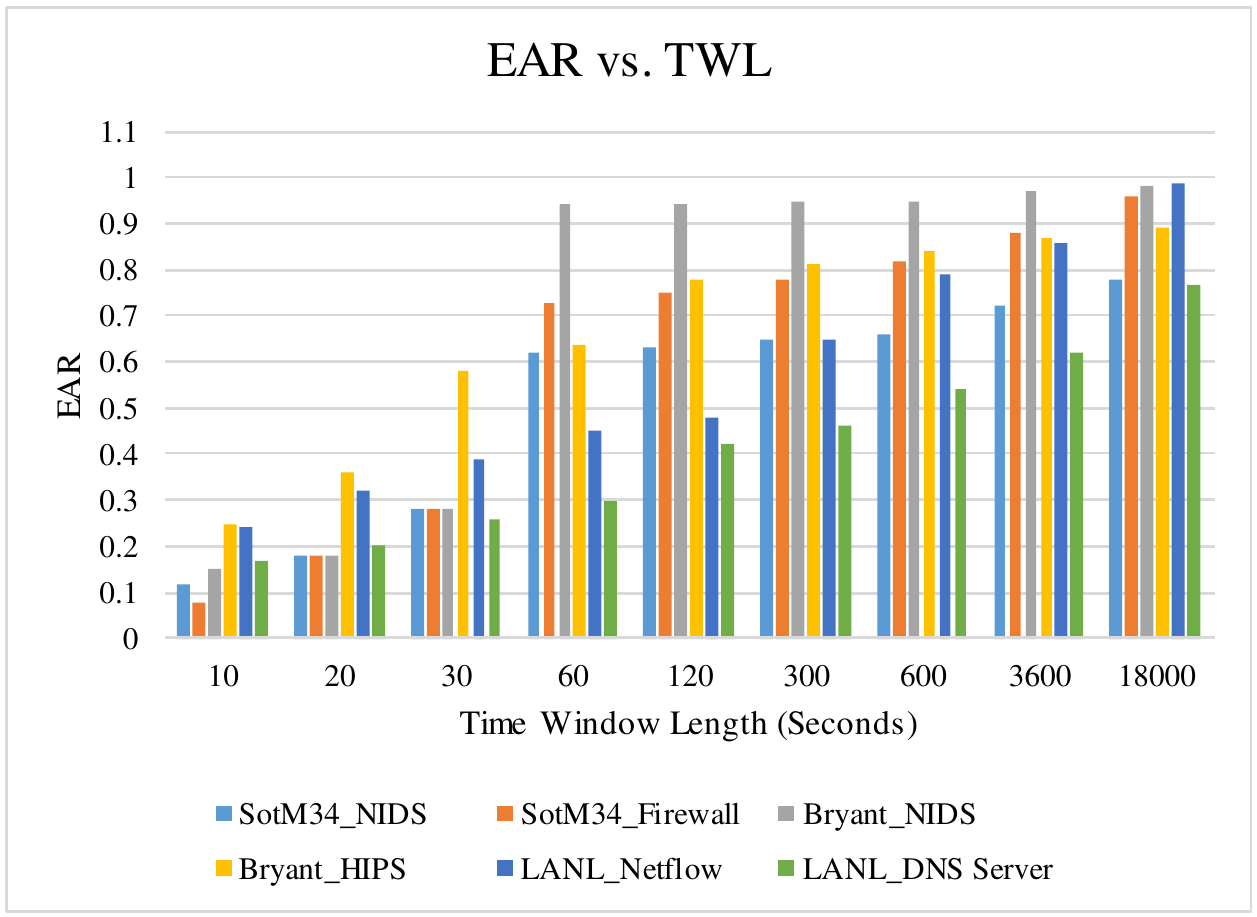}
\caption{The effect of TWL on the EAR metric}
\label{fig:10}       
\end{figure*}

During the execution of the proposed event aggregation method, one of the most effective parameters on the aggregation performance is the value of \textit{TWL} parameter for the aggregation and summarization analysis. According to the results of the experiments on the various types of events in the three different datasets (Fig.~\ref{fig:10}), it can be inferred that the size of this parameter has a direct impact on the EAR metric. Regarding the results, the longer the \textit{TWL} is, the higher the EAR will be. This analysis applies to all event types of different detection levels as shown in the figure. It should be noted that Network-level sensors have a higher EAR value compared to the sensors of the other detection levels, because of their higher event generation rates with a less time interval for similar logged events.

\subsubsection{The Relationship between EPR and TWL}
\label{4.2.2.epr-twl}
\begin{figure*}[ht]
\captionsetup{justification=centering}
\centering
  \includegraphics[width=0.7\textwidth]{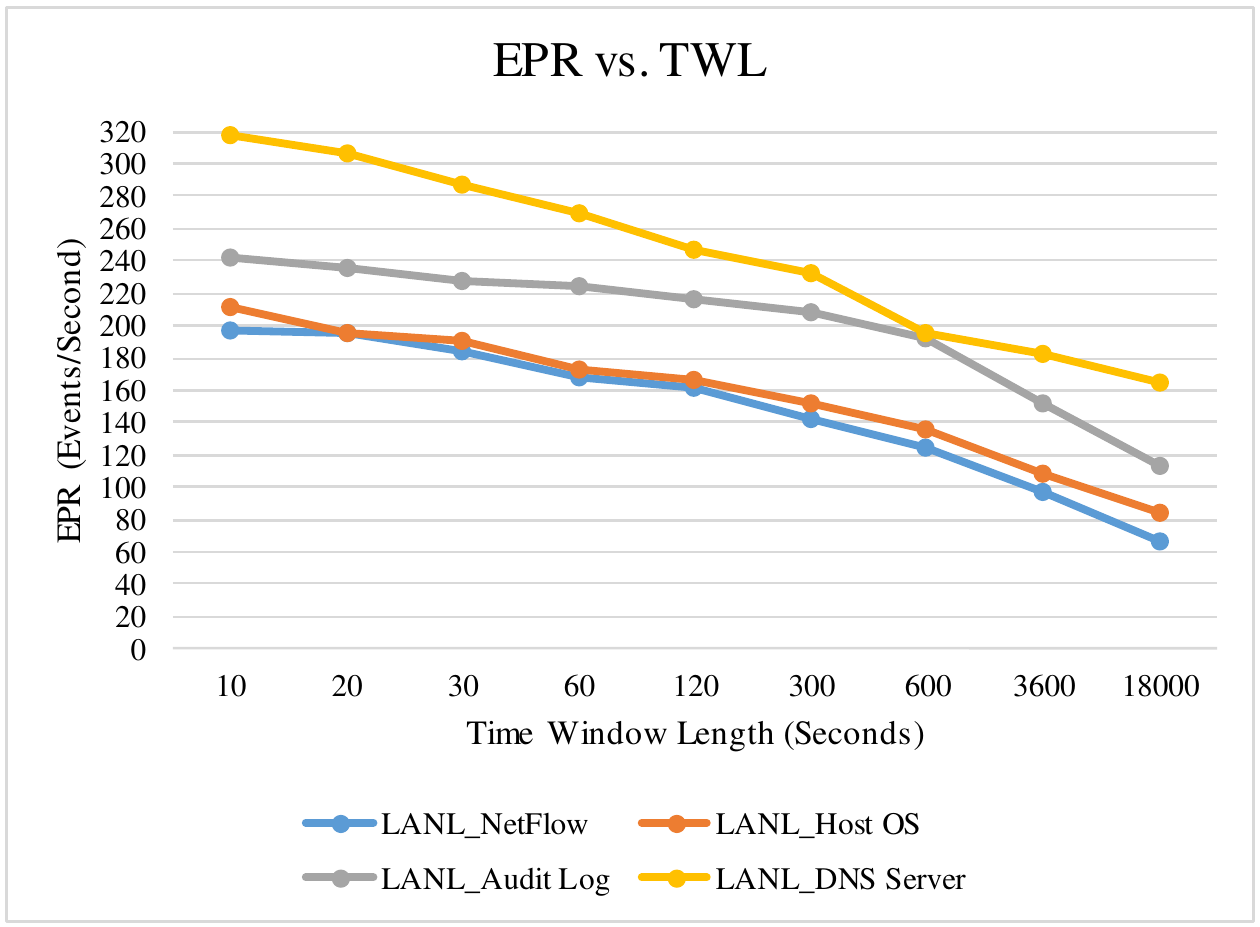}
\caption{The effect of TWL on the EPR metric}
\label{fig:11}       
\end{figure*}

The second analysis is investigating the effect of the \textit{TWL} on the execution time of the event aggregation algorithms and thus examining the rate of EPR. Fig.~\ref{fig:11} shows the results of algorithm execution on the LANL dataset's events with different values for the \textit{TWL} parameter. The results of the experiments show that for all types of events if the \textit{TWL} is set to a smaller value, the event aggregation, event filtration, and event summarization tasks (as core activities of the method) are faster due to fewer comparisons and computations. In addition, the number of features in \textit{NSFS} and \textit{SFS} for an event type affects overall performance. For example, based on Fig.~\ref{fig:11}, events of DNS Server are faster processed than NetFlow Sensor events due to having fewer features. 

\subsubsection{Aggregation Performance Curve (APC)}
\label{4.2.4.stor}
\begin{figure*}[ht]
\captionsetup{justification=centering}
\centering
  \includegraphics[width=0.7\textwidth]{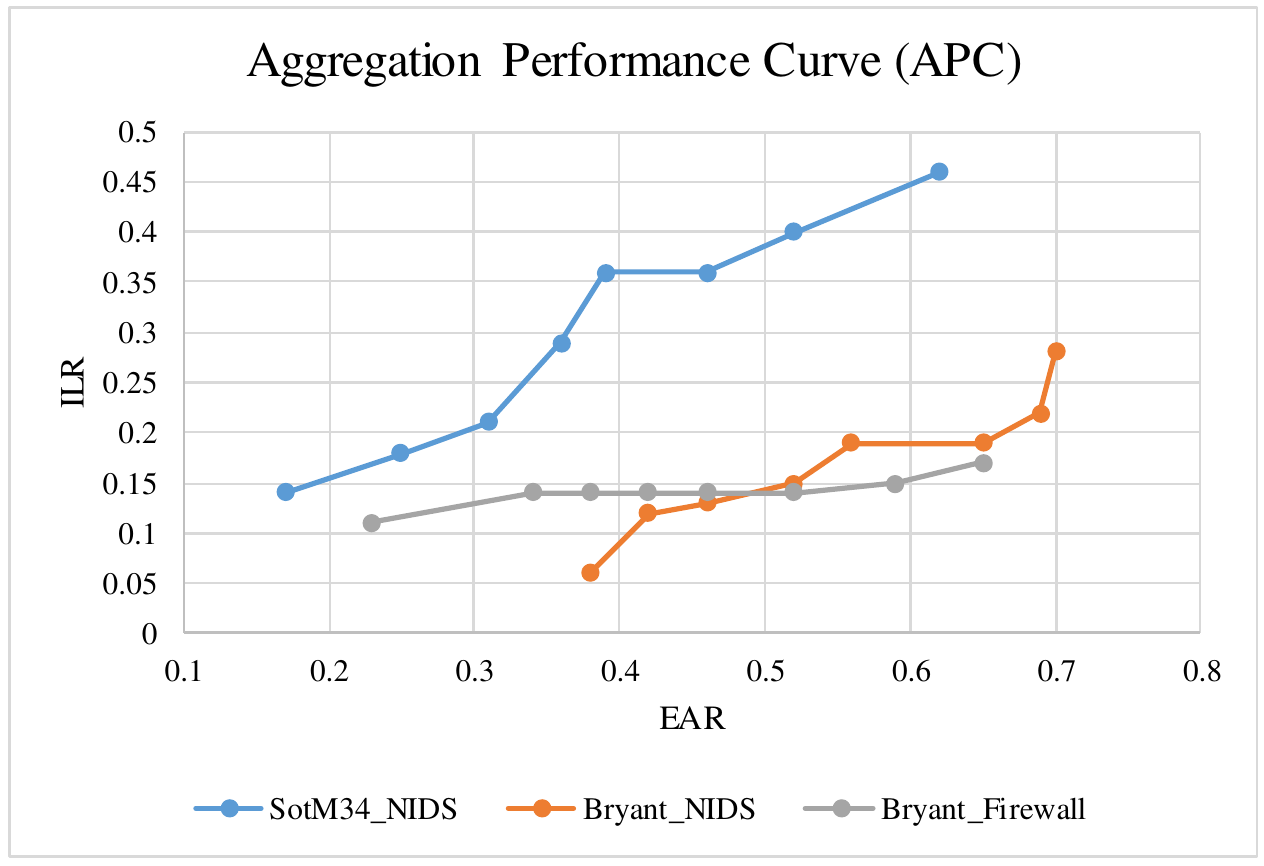}
\caption{The APC for the NIDS and Firewall sensors in our experiments}
\label{fig:13}       
\end{figure*}

The other main analysis in the area of intrusion event aggregation is drawing the aggregation performance curve (APC) \cite{r46}. The APC presents the relationships between the EAR and ILR metrics when threshold vectors of the \textit{SF} set vary. In this experiment, the NIDS events from the SotM34 dataset and the NIDS and Firewall events from the Bryant dataset are used for the evaluation due to having similar event features. According to the experiment, the aggregation method is repeated 8 times based on the 8 different threshold vectors for the \textit{SF} set of the events (Table~\ref{tbl:13}). In each iteration of the algorithm, the EAR and ILR metrics are computed for each identical threshold vector. The result of the experiment is shown in Fig.~\ref{fig:13} for all the three mentioned sensor events (for each curve, the leftmost point is for V1 and the rightmost point is for V8). Definable threshold vectors in the event summarization component allow the security manager to adjust the threshold values of each feature of different sensors based on the amount of summary required. In other words, based on this capability, a flexible way to summarize the events can be provided.

\begin{table}[ht]
\centering
\captionsetup{justification=centering}
\caption{The threshold vectors for the \textit{SFS} of the NIDS and Firewall sensors in our experiments} \label{tbl:13}
\vspace*{-4mm}
\includegraphics[width=0.8\textwidth]{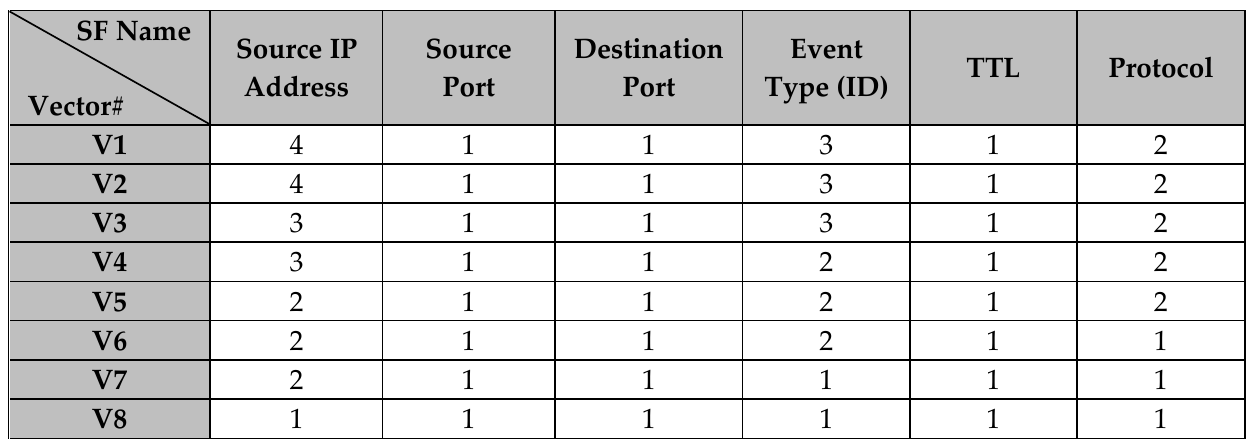}
\end{table}

According to the results of the experiment, it can be inferred that the EAR and ILR metrics have a direct impact on each other. In other words, when the value of the EAR metric increases, the value of ILR increases too. In addition, changes in the values of the threshold vector are also effective on the EAR and ILR values. It means that changing from a threshold vector to another one may cause a notable change in EAR and ILR behavior or not. For example, changes in the values of the threshold vector, despite the increase in EAR metric for the Firewall events, it does not have much influence on the ILR metric due to the nature of event feature values. Therefore, the efficiency of the aggregation method cannot be judged from the EAR alone, but its impact on the ILR, which is very important, must also be considered by the security analysts.

\subsubsection{The Effect of the TWL on Storage Space Reduction}
\label{4.2.3.stor}
\begin{figure*}[ht]
\captionsetup{justification=centering}
\centering
  \includegraphics[width=0.7\textwidth]{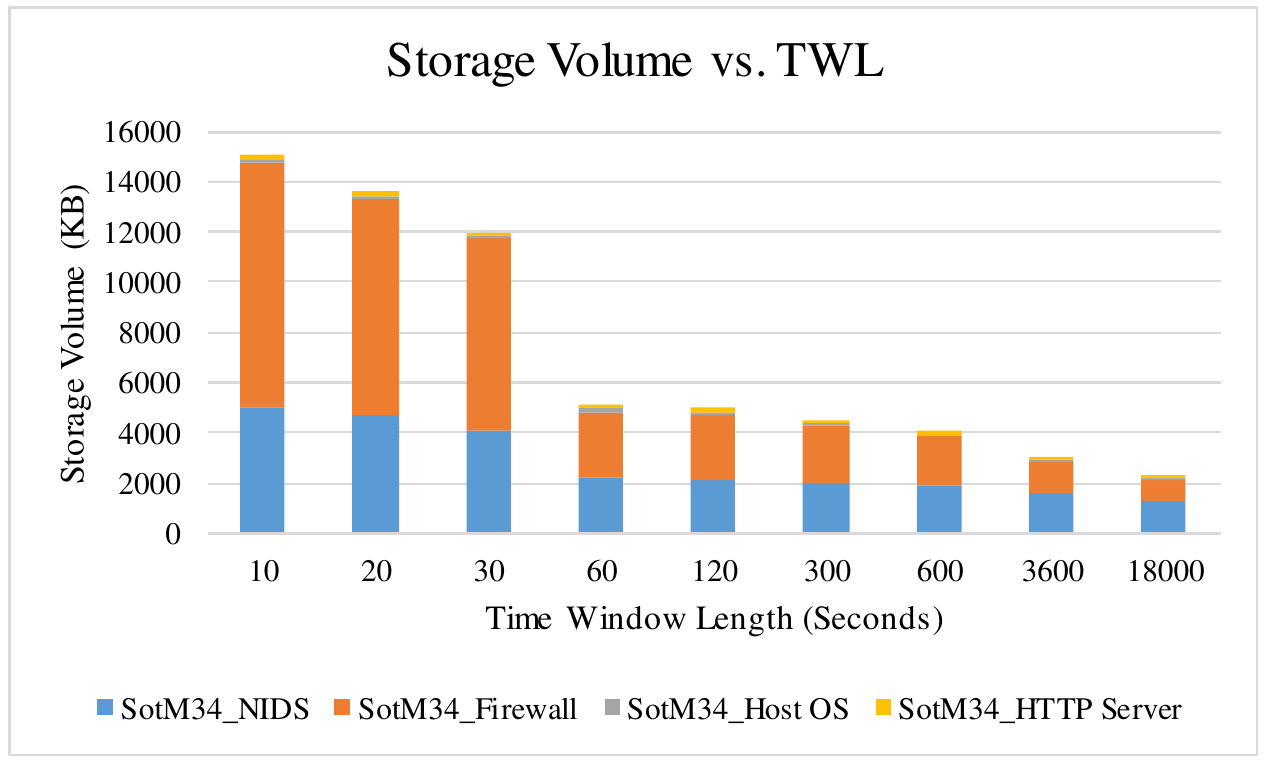}
\caption{The effect of the \textit{TWL} on storage space reduction}
\label{fig:12}       
\end{figure*}

The third analysis relates to the study of the effect of \textit{TWL} parameter on required storage space in real applications i.e. SIEM and SOC, which is important for security practitioners. In the mentioned systems, the storage space is always one of the main challenges to store the low-level events for the rest analysis i.e. attack strategy mining and forensics investigations. Regarding the proposed aggregation method, the effect of different \textit{TWL} values on the needed storage is illustrated in Fig.~\ref{fig:12} by using the SotM34 dataset events. Based on the results, the longer the \textit{TWL} are adjusted, the less storage space is required to store output summarized events. For example, for the Firewall sensor, if the \textit{TWL} is set to 18000 seconds (5 hours), the storage capacity is reduced by about 6 times when the \textit{TWL} is set to 30 seconds. Generally, it is recommended to set a higher \textit{TWL} where the rate of event generation per second is high by the sensor and the storage capacity is low.

\subsubsection{Improving Time to Detect and Respond}
\label{4.2.5.graph}
The other advantage that the proposed method provides is improving the time to detect and respond to security incidents. According to the literature, there are a lot of techniques for aggregated events post-processing i.e. correlation analysis to mine complex attack strategies and then fire appropriate responses to counter them \cite{r12}\cite{r15}\cite{r35}. One of the main event correlation techniques is graph-based attack analysis. In these techniques, aggregated events are modeled into a directed or undirected graph called an event correlation graph (ECG). In these approaches, ECG nodes are events. If two single events are at least similar in terms of the value of a field, then there will be an edge between the two. In addition, a weighted version of the ECG can be constructed in which the weight of ECG edges represents the correlation strength between logged events. One of the main challenges for the graph-based event correlation techniques is the graph complexity with the higher number of nodes and edges. For example, some related works such as \cite{r12}\cite{r43}\cite{r69}\cite{r70} suffer from this limitation. Hence, event aggregation and summarization analyses can help them to minimize the complexity of the final generated graph model (with less number of nodes and edges) which is used for later processing i.e. detecting various stages of an IKC.

\begin{figure*}[ht]
\captionsetup{justification=centering}
\centering
  \includegraphics[width=0.9\textwidth]{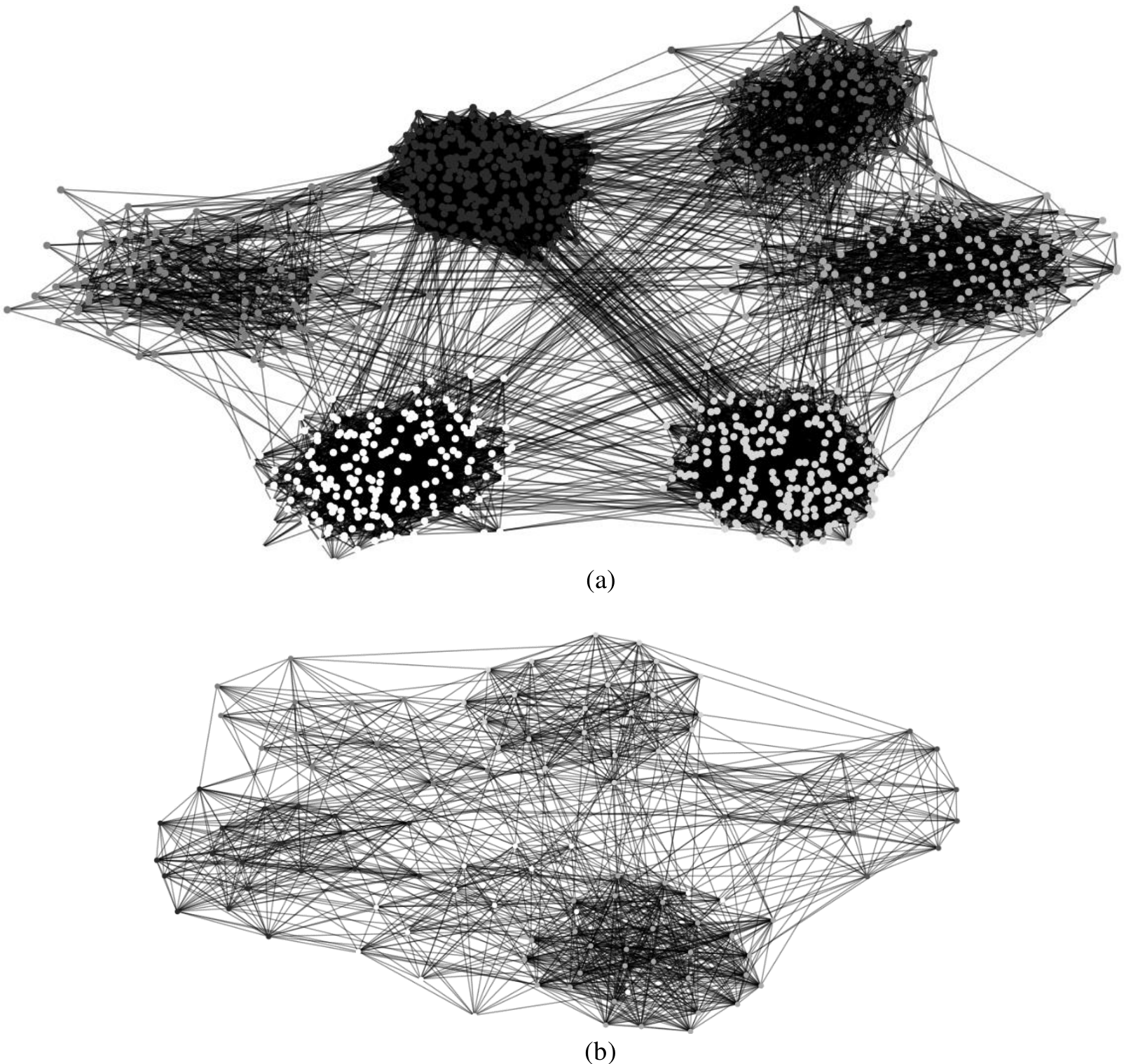}
\caption{Sample constructed ECGs, (a) with low-level events and (b) with aggregated events}
\label{fig:14}       
\end{figure*}

For example, suppose that the event correlation system constructs an ECG from the logged events to discover the IKC stages based on the graph analysis and finally reconstruct the APT attack scenario. This correlation can be based on the existence of features with common values in the two distinct events, according to which an edge between them is inserted in the constructed ECG. Two sample ECGs for a set of events of the Bryant dataset with low-level events (part (a)) and with aggregated events (part (b)) are provided in Fig.~\ref{fig:14}. According to the ECG, if we want to extract the connection of an external machine (attacker) to a node of the internal network (victim) for detecting \textit{Delivery} stage of IKC based on graph analysis, then we must have an edge between two distinct nodes of the ECG, $e_1$ and $e_2$, where \{\{$e_1.sourceIP$ $\notin$ \{Internal IP Range $\lor$ Trusted IPs\} $\land$ \{$e_1.destinationIP$ $\in$ Internal IP range\} $\land$ \{$e_1.eventType$ $\in$ Reconnaissance Activity\}\} and \{\{$e_2.sourceIP$ $\notin$ \{Internal IP Range $\lor$ Trusted IPs\} $\land$ \{$e_2.destinationIP$ $\in$ Internal IP range\} $\land$ \{$e_2.eventType$ $\in$ Connection Receive $\lor$ File Create\}\} such that \{$e_1.sourceIP$=$e_2.sourceIP$ $\land$ $e_1.destinationIP$=$e_2.destinationIP$\}. It is obvious that finding such relationships between events in an ECG with less complexity (part (b) of Fig.~\ref{fig:14}), graph resulting from aggregated events with fewer nodes and edges, requires much less processing time for reducing mean time to detect and respond for security incidents. The results of comparison between the detection analysis of the two mentioned cases in Fig.~\ref{fig:14} is presented in Table~\ref{tbl:14}. Based on our examination, the execution time of detection analysis for the two graphs in Fig.~\ref{fig:14} is 12208 seconds and 1217 seconds for the constructed ECG with low-level events and the constructed ECG with aggregated events, respectively. As it can be seen, the event aggregation analysis has accelerated the discovery of the resulting IKC stages in the ECG by about tenfold. 

\begin{table}[ht]
\centering
\captionsetup{justification=centering}
\caption{The results of comparison between the detection analysis of the two mentioned cases in Fig.~\ref{fig:14}}\label{tbl:14}
\vspace*{-4mm}
\includegraphics[width=0.8\textwidth]{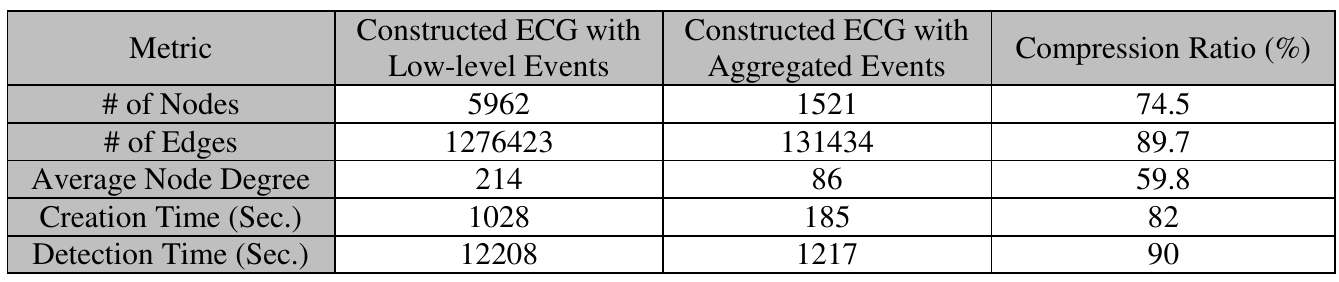}
\end{table}

\section{Conclusion and Future Work}\label{5:conc}
One of the main promising approaches to track the attacker's behaviors and detect the malicious activities during the targeted multi-stage attacks like APTs is the use of various heterogeneous security and non-security sensors in different lines of defense in a monitored network (Network, Host, and Application). One of the main challenges of this approach is the huge amount of logged events raised by heterogeneous sensors for tracking the malicious attackers behind APT attacks. The main objective of this paper is to propose an event aggregation method to reduce the volume of logged events by heterogeneous sensors. These low-level events are generated during the different attack stages of an IKC model of complex targeted cybersecurity attacks e.g. APTs. Such event aggregation analysis is an important necessity to ease the management of the events for later usages i.e. event correlation analysis for timely detecting of APT attack scenarios. The input of the proposed method is a set of events that are logged by a set of heterogeneous sensors which can be deployed in the three main detection levels of a target network, Network, Host, and Application. The proposed aggregation method has three main components as follows: 1) the event aggregation component for clustering similar logged events of each sensor based on an attribute-based similarity matching regarding some non-summarizable features of the sensor type (\textit{NSF} set), 2) the event filtration component for eliminating the noisy events from the output event clusters by using a clustering-based local outlier factor, and 3) the event summarization component for fusing remaining events and improving their quality by leveraging an attribute-oriented induction method regarding some summarizable features (\textit{SF} feature set) of each sensor type. Our implementation and experimental results have proved that the proposed method makes it possible to summarize heterogeneous events by eliminating redundant and false information with an acceptable level for the performance metrics.

In the future, some improvements can be made to the proposed aggregation method as follows: 1) One of the flaws of the proposed event aggregation method is the lack of a proper degree of scalability due to having a centralized architecture for aggregating and summarizing logged events. Data-intensive computing (also known as Big Data Analytics) is a promising technology to solve scalability by distributed processing of heterogeneous events, 2) The aggregation component of the proposed method works based on a constant \textit{TWL} for each of the sensors which may not be adequately efficient for environments where the attacker's behavior changes dynamically during attacks. To solve this problem, setting a dynamic time window length for the sensors based on predicting how long an attacker will stay in a unique IKC stage of APT attack will be a good solution, and 3) Another improvement that can be made to the event aggregation process is the ability to cluster the events hierarchically. It seems that by using this type of clustering analysis, the aggregation component can cluster events based on different levels of granularity which makes it easier to analyze the clustering output for the network security manager. The other main future work is proposing a general event correlation framework to detect APT attack scenarios based on the IKC models by analyzing aggregated events produced by the proposed aggregation method.

\section*{Appendix}
\label{6.appendix}
\begin{landscape}
\vspace*{3mm}
\begin{figure*}[ht]
\centering
\captionsetup{justification=centering}
\caption{Candidate \textit{NSFS} and \textit{SFS} of each heterogeneous security and non-security sensor in different detection levels (Network, Host, and Application)} \label{fig:6}
\vspace*{-4mm}
\includegraphics[width=1.5\textwidth,height=0.8\textheight]{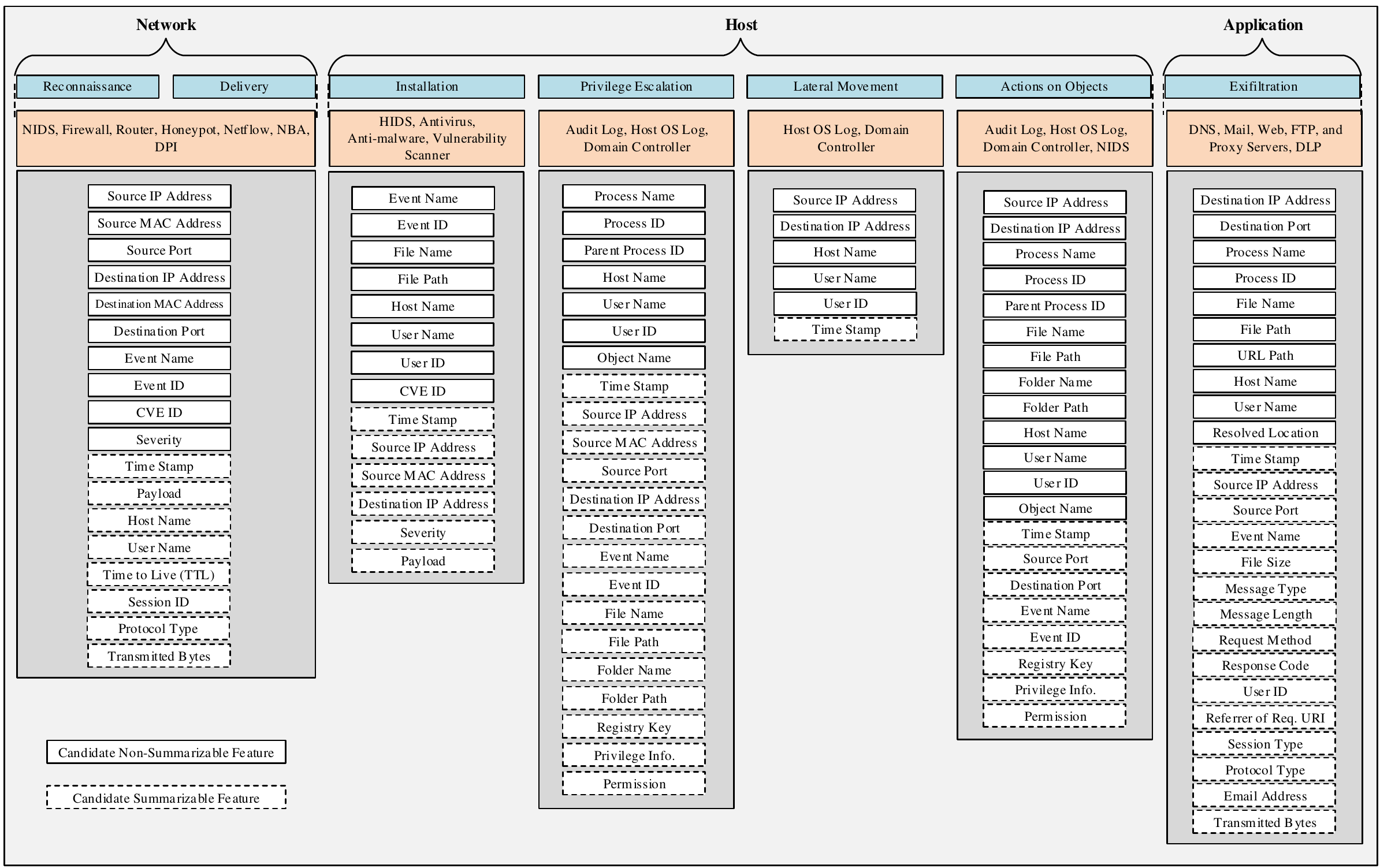}
\end{figure*}
\end{landscape}

\begin{table}[ht]
\centering
\captionsetup{justification=centering}
\caption{Details of the existing publicly available datasets in the intrusion detection field} \label{tbl:9}
\vspace*{-4mm}
\includegraphics[width=1\textwidth]{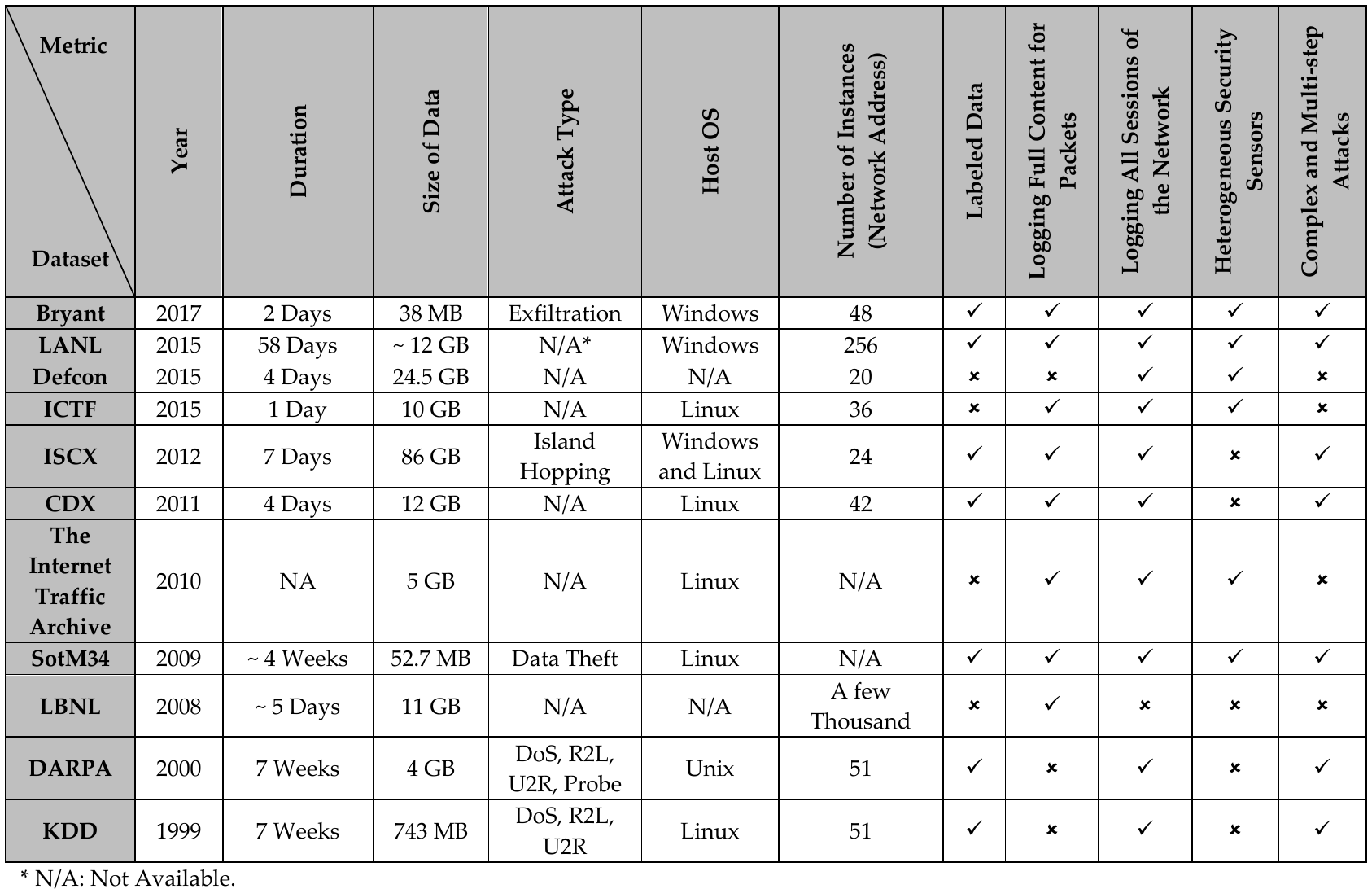}
\end{table} 

\begin{table}[ht]
\centering
\captionsetup{justification=centering}
\caption{The threshold vectors for \textit{SF} features of the sensors in our experiments} \label{tbl:10}
\vspace*{-4mm}
\includegraphics[width=0.8\textwidth]{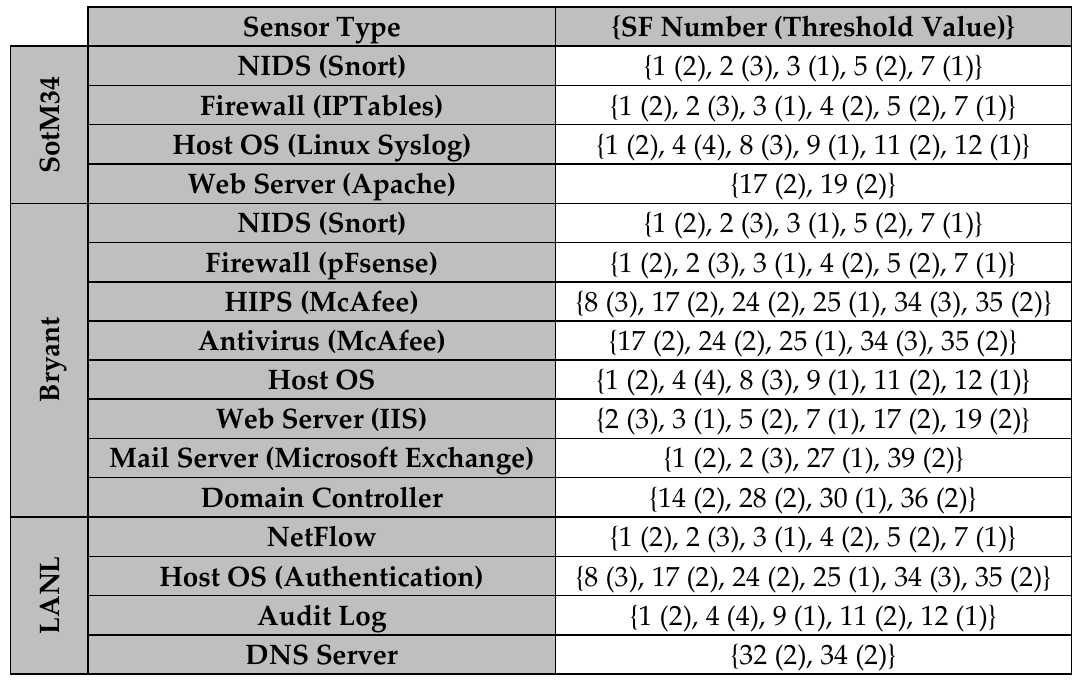}
\end{table}


\begin{thebibliography}{}
\bibitem{r1}Quintero-Bonilla, S., \& Martín del Rey, A. (2020). A new proposal on the advanced persistent threat: a survey. Applied Sciences, 10(11), 3874.
\bibitem{r2}Husak, M., Komarkova, J., Bou-Harb, E., \& Celeda, P. (2018). Survey of Attack Projection, Prediction, and Forecasting in Cyber Security. IEEE Communications Surveys \& Tutorials.
\bibitem{r3}Singh, S., Sharma, P. K., Moon, S. Y., Moon, D., \& Park, J. H. (2019). A comprehensive study on APT attacks and countermeasures for future networks and communications: challenges and solutions. The Journal of Supercomputing, 75(8), 4543-4574.
\bibitem{r4}Kim, G., Lee, C., Jo, J., \& Lim, H. (2020). Automatic extraction of named entities of cyber threats using a deep Bi-LSTM-CRF network. International Journal of Machine Learning and Cybernetics, 11(10), 2341-2355.
\bibitem{r5}Khosravi-Farmad, M., Ramaki, A. A., \& Bafghi, A. G. (2018, October). Moving Target Defense against Advanced Persistent Threats for Cybersecurity Enhancement. In 2018 8th International Conference on Computer and Knowledge Engineering (ICCKE) (pp. 280-285). IEEE.
\bibitem{r6}Bahrami, P. N., Dehghantanha, A., Dargahi, T., Parizi, R. M., Choo, K. K. R., Javadi, H. H., ... \& Ren, W. (2019). A layered security architecture based on cyber kill chain against advanced persistent threats.
\bibitem{r7}Hutchins, E. M., Cloppert, M. J., \& Amin, R. M. (2011). Intelligence-driven computer network defense informed by analysis of adversary campaigns and intrusion kill chains. Leading Issues in Information Warfare \& Security Research, 1(1), 80.
\bibitem{r8}Luh, R., Marschalek, S., Kaiser, M., Janicke, H., \& Schrittwieser, S. (2017). Semantics-aware detection of targeted attacks: a survey. Journal of Computer Virology and Hacking Techniques, 13(1), 47-85.
\bibitem{r9}Bryant, B. D., \& Saiedian, H. (2017). A novel kill-chain framework for remote security log analysis with SIEM software. Computers \& Security, 67, 198-210.
\bibitem{r10}Wilkens, F., Ortmann, F., Haas, S., Vallentin, M., \& Fischer, M. (2021). Multi-Stage Attack Detection via Kill Chain State Machines. arXiv preprint arXiv:2103.14628.
\bibitem{r11}Spadaro, A. (2013). Event correlation for detecting advanced multi-stage cyber-attacks (Doctoral dissertation, Delft University of Technology).
\bibitem{r12}Pei, K., Gu, Z., Saltaformaggio, B., Ma, S., Wang, F., Zhang, Z., ... \& Xu, D. (2016, December). Hercule: Attack story reconstruction via community discovery on correlated log graph. In Proceedings of the 32Nd Annual Conference on Computer Security Applications (pp. 583-595). ACM.
\bibitem{r13}Salah, S., Maciá-Fernández, G., \& DiAz-Verdejo, J. E. (2013). A model-based survey of alert correlation techniques. Computer Networks, 57(5), 1289-1317.
\bibitem{r14}Syed, R. H., Pazardzievska, J., \& Bourgeois, J. (2012). Fast attack detection using correlation and summarizing of security alerts in grid computing networks. The Journal of Supercomputing, 62(2), 804-827.
\bibitem{r15}Ramaki, A. A., Rasoolzadegan, A., \& Bafghi, A. G. (2018). A systematic mapping study on intrusion alert analysis in intrusion detection systems. ACM Computing Surveys (CSUR), 51(3), 55.
\bibitem{r16}Valeur, F., Vigna, G., Kruegel, C., \& Kemmerer, R. A. (2004). Comprehensive approach to intrusion detection alert correlation. IEEE Transactions on dependable and secure computing, 1(3), 146-169.
\bibitem{r17}Kim, H., Kwon, H., \& Kim, K. K. (2019). Modified cyber kill chain model for multimedia service environments. Multimedia Tools and Applications, 78(3), 3153-3170.
\bibitem{r18}Rutherford, J. R., \& White, G. B. (2016, January). Using an improved cybersecurity kill chain to develop an improved honey community. In System Sciences (HICSS), 2016 49th Hawaii International Conference on (pp. 2624-2632). IEEE.
\bibitem{r19}Pandey, S. K., \& Mehtre, B. M. (2014, April). A lifecycle based approach for malware analysis. In Communication Systems and Network Technologies (CSNT), 2014 Fourth International Conference on (pp. 767-771). IEEE.
\bibitem{r20}Center, M. I. (2013). APT1: Exposing one of China's cyber espionage units. Mandiant. com.
\bibitem{r21}Flynn, J. (2012). Intrusion along the kill chain. Proceedings of BlackHat USA. BlackHat.
\bibitem{r22}Advanced Persistent Threats: A Decade in Review, Command Five Pty Ltd., 2011. Available:\href{http://www.commandfive.com/papers/C5_APT_ADecadeInReview.pdf}.
\bibitem{r23}Dell SecureWorks. (2015). Breaking the kill chain- knowing, detecting, disrupting and eradicating the advanced threat. Available: \href{https://webobjects.cdw.com/webobjects/media/pdf/paloalto/Breaking-the-Attack-Kill-Chain.pdf}.
\bibitem{r24}Barnum, S. (2012). Standardizing cyber threat intelligence information with the Structured Threat Information eXpression (STIX). MITRE Corporation, 11, 1-22.
\bibitem{r25}Lima, A. J. C. (2015). Advanced persistent threats (Doctoral dissertation).
\bibitem{r26}Hudson, B. (2014). Advanced Persistent Threats: Detection, Protection and Prevention. Sophos Ltd., US February.
\bibitem{r27}Tonelli, E. (2014). WatchGuard APT Blocker - WatchGuard Platinum Partner, WatchGuard Technologies, Inc.
\bibitem{r28}Alex Cox. Stalking the kill chain: The attacker's chain. Available: \href{http://blogs.rsa.com/stalking-the-kill-chain-the-attackers-chain-2/}, 2012.
\bibitem{r29}Chen, P., Desmet, L., \& Huygens, C. (2014, September). A study on advanced persistent threats. In IFIP International Conference on Communications and Multimedia Security (pp. 63-72). Springer, Berlin, Heidelberg.
\bibitem{r30}Websense. Advanced Persistent Threats and other Advanced Attacks. 2013. Available: \href{https://www.websense.com/assets/html/apt/apt-overview-from-fud-to-facts.pdf}.
\bibitem{r31}Bhatt, P., Yano, E. T., \& Gustavsson, P. (2014, April). Towards a framework to detect multi-stage advanced persistent threats attacks. In Service Oriented System Engineering (SOSE), 2014 IEEE 8th International Symposium on (pp. 390-395). IEEE.
\bibitem{r32}Giura, P., \& Wang, W. (2013). Using large scale distributed computing to unveil advanced persistent threats. Science, 1(3), 93-102.
\bibitem{r33}Bhatt, S., Manadhata, P. K., \& Zomlot, L. (2014). The operational role of security information and event management systems. IEEE Security \& Privacy, 12(5), 35-41.
\bibitem{r34}Giura, P., \& Wang, W. (2012, December). A context-based detection framework for advanced persistent threats. In Cyber Security (CyberSecurity), 2012 International Conference on (pp. 69-74). IEEE.
\bibitem{r35}Ramaki, A. A., Amini, M., \& Atani, R. E. (2015). RTECA: Real time episode correlation algorithm for multi-step attack scenarios detection. Computers \& Security, 49, 206-219.
\bibitem{r36}Husak, M., Cermak, M., Lastovicka, M., \& Vykopal, J. (2017, May). Exchanging security events: Which and how many alerts can we aggregate?. In 2017 IFIP/IEEE Symposium on Integrated Network and Service Management (IM) (pp. 604-607). IEEE.
\bibitem{r37}Spathoulas, G. P., \& Katsikas, S. K. (2013). Enhancing IDS performance through comprehensive alert post-processing. Computers \& Security, 37, 176-196.
\bibitem{r38}Fredj, O. B. (2015). A realistic graph-based alert correlation system. Security and Communication Networks, 8(15), 2477-2493.
\bibitem{r39}Sun, Y., \& Chen, X. (2020). An Improved Frequent Pattern Growth Based Approach to Intrusion Detection System Alert Aggregation. In Journal of Physics: Conference Series (Vol. 1437, No. 1, p. 012070). IOP Publishing.
\bibitem{r40}Alserhani, F. M. (2016). Alert correlation and aggregation techniques for reduction of security alerts and detection of multistage attack. International Journal of Advanced Studies in Computers, Science and Engineering, 5(2), 1.
\bibitem{r41}Shittu, R., Healing, A., Ghanea-Hercock, R., Bloomfield, R., \& Rajarajan, M. (2015). Intrusion alert prioritisation and attack detection using post-correlation analysis. Computers \& Security, 50, 1-15.
\bibitem{r42}Soleimani, M., \& Ghorbani, A. A. (2012). Multi-layer episode filtering for the multi-step attack detection. Computer Communications, 35(11), 1368-1379.
\bibitem{r43}de Alvarenga, S. C., Barbon Jr, S., Miani, R. S., Cukier, M., \& Zarpelão, B. B. (2018). Process mining and hierarchical clustering to help intrusion alert visualization. Computers \& Security, 73, 474-491.
\bibitem{r44}Zhang, R., Guo, T., \& Liu, J. (2018, March). An IDS Alerts Aggregation Algorithm Based on Rough Set Theory. In IOP Conference Series: Materials Science and Engineering (Vol. 322, No. 6, p. 062009). IOP Publishing.
\bibitem{r45}Kim, J., Moon, I., Lee, K., Suh, S. C., \& Kim, I. (2015, March). Scalable security event aggregation for situation analysis. In Big Data Computing Service and Applications (BigDataService), 2015 IEEE First International Conference on (pp. 14-23). IEEE.
\bibitem{r46}Saad, S., \& Traore, I. (2012). Heterogeneous Multi-sensor IDS Alerts Aggregation using Semantic Analysis. Journal of Information Assurance \& Security, 7(2).
\bibitem{r47}Ahmadian Ramaki, A., \& Rasoolzadegan, A. (2016). Causal knowledge analysis for detecting and modeling multi-step attacks. Security and Communication Networks, 9(18), 6042-6065.
\bibitem{r48}Nadeem, A., Verwer, S., Moskal, S., \& Yang, S. J. (2021). SAGE: Intrusion Alert-driven Attack Graph Extractor. arXiv preprint arXiv:2107.02783.
\bibitem{r49}Ahmed, M. (2018). Data summarization: a survey. Knowledge and Information Systems, 1-25.
\bibitem{r50}Jiang, J., Chen, J., Choo, K. K. R., Liu, C., Liu, K., \& Yu, M. (2017, October). A Visualization Scheme for Network Forensics Based on Attribute Oriented Induction Based Frequent Item Mining and Hyper Graph. In International Conference on Digital Forensics and Cyber Crime (pp. 130-143). Springer, Cham.
\bibitem{r74}Vega, C., Roquero, P., Leira, R., Gonzalez, I., \& Aracil, J. (2017). Loginson: a transform and load system for very large-scale log analysis in large IT infrastructures. The Journal of Supercomputing, 73(9), 3879-3900.
\bibitem{r51}Chhajed, S. (2015). Learning ELK stack. Packt Publishing Ltd.
\bibitem{r52}Negoita, O., \& Carabas, M. (2020, July). Enhanced Security Using Elasticsearch and Machine Learning. In Science and Information Conference (pp. 244-254). Springer, Cham.
\bibitem{r53}Sapegin, A., Jaeger, D., Azodi, A., Gawron, M., Cheng, F., \& Meinel, C. (2013, December). Hierarchical object log format for normalization of security events. In Information Assurance and Security (IAS), 2013 9th International Conference on (pp. 25-30). IEEE.
\bibitem{r54}Zhao, X., Liang, J., \& Cao, F. (2014). A simple and effective outlier detection algorithm for categorical data. International Journal of Machine Learning and Cybernetics, 5(3), 469-477.
\bibitem{r55}Amer, M., \& Goldstein, M. (2012, August). Nearest-neighbor and clustering based anomaly detection algorithms for Rapidminer. In Proc. of the 3rd RapidMiner Community Meeting and Conference (RCOMM 2012) (pp. 1-12).
\bibitem{r56}He, Z., Xu, X., \& Deng, S. (2003). Discovering cluster-based local outliers. Pattern Recognition Letters, 24(9-10), 1641-1650.
\bibitem{r64}Chuvakin A. Scan of the Month 34. Available: \href{http://www.honeynet.org/scans/scan34/}.
\bibitem{r65}Kent, A. D. (2015). Comprehensive, multi-source cyber-security events data set (No. LA-UR-15-23810). Los Alamos National Lab (LANL), Los Alamos, NM (United States). Available: \href{https://csr.lanl.gov/data/cyber1/}.
\bibitem{r66}Halkidi, M., Batistakis, Y., \& Vazirgiannis, M. (2002). Clustering validity checking methods: part II. ACM Sigmod Record, 31(3), 19-27.
\bibitem{r67}Dziopa, T. (2016, January). Clustering Validity Indices Evaluation with Regard to Semantic Homogeneity. In FedCSIS Position Papers (pp. 3-9).
\bibitem{r68}Hassanzadeh, A., \& Burkett, R. (2018, August). SAMIIT: Spiral attack model in IIoT mapping security alerts to attack life cycle phases. In 5th International Symposium for ICS \& SCADA Cyber Security Research 2018 5 (pp. 11-20).
\bibitem{r69}Debatty, T., Mees, W., \& Gilon, T. (2018, May). Graph-based APT detection. In 2018 International Conference on Military Communications and Information Systems (ICMCIS) (pp. 1-8). IEEE.
\bibitem{r70}Milajerdi, S. M., Gjomemo, R., Eshete, B., Sekar, R., \& Venkatakrishnan, V. N. (2019, May). Holmes: real-time apt detection through correlation of suspicious information flows. In 2019 IEEE Symposium on Security and Privacy (SP) (pp. 1137-1152). IEEE.
\end{thebibliography}
\end{document}